\documentclass[12pt]{article}
\usepackage{graphicx}
\usepackage{amssymb}
\usepackage{amsmath}
\usepackage{epsfig}

\usepackage{epsf}
\usepackage{graphicx,epsfig}
\usepackage{amsfonts}
\usepackage{amssymb}

\usepackage{euscript}
\usepackage{amssymb}
\usepackage{amsfonts}
\usepackage{amsbsy}
\usepackage{amsmath}
\usepackage{epsfig}
\usepackage{amsthm}
\usepackage{amscd}
\usepackage{amstext}
\usepackage{hyperref}

\newcommand{\vol}{{\rm{vol}}}
\newcommand{\mc}{\mathcal}
\renewcommand{\t}{\tilde}

\def\gsim{\, \rlap{$>$}{\lower 1.1ex\hbox{$\sim$}}\,}
\def\lsim{\, \rlap{$<$}{\lower 1.1ex\hbox{$\sim$}}\,}

\makeatletter
\renewcommand\section{\@startsection {section}{1}{\z@}%
                                 {-3.5ex \@plus -1ex \@minus -.2ex}
                                   {2.3ex \@plus.2ex}%
                                   {\normalfont\large\bfseries}}
\renewcommand\subsection{\@startsection{subsection}{2}{\z@}%
                                   {-3.25ex\@plus -1ex \@minus -.2ex}%
                                     {1.5ex \@plus .2ex}%
                                     {\normalfont\bfseries}}
\renewcommand\subsubsection{\@startsection{subsubsection}{3}{\z@}%
                                   {-3.25ex\@plus -1ex \@minus -.2ex}%
                                     {1.5ex \@plus .2ex}%
                                     {\normalfont\itshape}}
\makeatother

\pdfoutput=1

\newcommand{\be}{\begin{equation}}
\newcommand{\ee}{\end{equation}}
\newcommand{\bea}{\begin{eqnarray}}
\newcommand{\eea}{\end{eqnarray}}

\newcommand{\barr}{\begin{array}}
\newcommand{\earr}{\end{array}}

\def\beq{\begin{equation}}
\def\eeq{\end{equation}}
\def\be{\begin{equation}}
\def\ee{\end{equation}}
\def\bea{\begin{eqnarray}}
\def\eea{\end{eqnarray}}

\DeclareRobustCommand{\SkipTocEntry}[4]{}

\begin{document}

\begin{titlepage}

\setcounter{page}{1} \baselineskip=15.5pt \thispagestyle{empty}

\begin{flushright}
NSF-KITP-10-068\\
SU-ITP-10/19\\
SLAC-PUB-14146
\end{flushright}
\vfil

\begin{center}
{\LARGE  Micromanaging de Sitter holography}

\end{center}
\bigskip\

\begin{center}
{\large Xi Dong$^{2,1}$, Bart Horn$^{2,1}$, Eva Silverstein$^{1,2^*}$, Gonzalo Torroba$^{2,1}$}
\end{center}

\begin{center}
\textit{$^1$Kavli Institute for Theoretical Physics and Department of Physics, University of California, Santa Barbara CA 93106
}
\end{center}

\begin{center}
\textit{
$^2$SLAC and Department of Physics, Stanford University, Stanford CA 94305
}
\end{center} \vfil

\vspace{.8cm}

\hrule \vspace{0.3cm}
{\small  \noindent \textbf{Abstract} \\[0.3cm]
\noindent
We develop tools to engineer de Sitter vacua with semi-holographic duals, using elliptic fibrations and orientifolds to uplift Freund-Rubin compactifications with CFT duals.  The dual brane construction is compact and constitutes a microscopic realization of the dS/dS correspondence, realizing d-dimensional de Sitter space as a warped compactification down to ($d-1$)-dimensional de Sitter gravity coupled to a pair of large-N matter sectors.
This provides a parametric microscopic interpretation of the Gibbons-Hawking entropy.  We illustrate these ideas with an explicit class of examples in three dimensions, and describe ongoing work on four-dimensional constructions.
}\vspace{0.5cm}  \hrule

\noindent

\vfil
\begin{flushleft}
\today
\end{flushleft}

$^*$ On leave

\end{titlepage}

\newpage
\tableofcontents
\newpage

\section{Introduction}

The Gibbons-Hawking entropy of the de Sitter horizon \cite{Gibbons:1977mu}\ invites a microscopic interpretation and a holographic formulation of inflating spacetimes.  Much progress was made in the analogous problem in black hole physics using special black holes in string theory whose microstates could be reliably counted, such as those analyzed in \cite{Strominger:1996sh,Callan:1996dv}; this led to the AdS/CFT correspondence \cite{AdS/CFT}. In contrast, a microscopic understanding of the entropy of de Sitter space is more difficult for several reasons including its potential dynamical connections to other backgrounds (metastability), the absence of a non-fluctuating timelike boundary, and the absence of supersymmetry.

In this paper, we develop a class of de Sitter constructions in string theory, built up from AdS/CFT dual pairs along the lines of \cite{Polchinski:2009ch}, which are simple enough to provide a microscopic accounting of the parametric scaling of the Gibbons-Hawking entropy.
These models realize microscopically a semi-holographic description of metastable de Sitter space which had been derived macroscopically in \cite{Alishahiha:2004md}.  It would also be interesting to connect this to other approaches to de Sitter holography such as \cite{dSCFT,hats}\ and to other manifestations of the de Sitter entropy such as \cite{SeeingS}.\footnote{See \cite{Banks}\ for a different proposal.}   The construction is somewhat analogous to neutral black branes analyzed in \cite{Danielsson:2001xe}.

We will begin in \S \ref{sec:holography} by explaining the salient features of the holographic duality and of the de Sitter construction which realizes it microscopically.  In \S \ref{sec:microscopic} we will lay out our methods in more detail, applying them to worked examples of $dS_3$ in \S \ref{sec:dSthree}.  Finally, \S \ref{sec:discussion} discusses further directions and ongoing work, including $dS_4$ constructions in progress.

\section{dS holography and microscopy}\label{sec:holography}

A semi-holographic duality follows simply by recognizing the de Sitter static patch as a warped compactification
\beq\label{dSdSmetric}
ds^2_{dS_d}=dw^2+\sin^2\left({w\over{R_{dS}}}\right)ds_{dS_{d-1}}^2.
\eeq
The warp factor $\sin^2(w/R_{dS})$ goes to zero at $w=0,\pi R_{dS}$ and rises to a finite maximum in between, implying two warped throats and a propagating graviton in $d-1$ dimensions.  Such a semi-holographic duality is familiar in the study of warped compactifications (such as \cite{Giddings:2001yu})  and Randall-Sundrum models \cite{Randall:1999vf}.  In these systems, the bulk of the throats admits a dual description in terms of a field theory (as in \cite{Klebanov:2000hb}), but the finite maximum of the warp factor implies that this field theory is cut off at a finite scale and coupled to gravity \cite{Randall:1999vf,Gubser:1999vj,Hawking:2000da}.
The main observation in \cite{Alishahiha:2004md}\ was that the same statements apply to de Sitter (\ref{dSdSmetric}).

This macroscopic derivation of a holographic description leaves open the question of what degrees of freedom build up the two throats microscopically.
In this work, we find that `uplifting' AdS/CFT brane constructions to de Sitter space automatically produces the two-throat structure, while revealing (example by example) the microscopic degrees of freedom that build up the throats.

Before turning to the detailed examples, let us explain the main features of the construction and its realization of de Sitter holography.
Freund-Rubin solutions of the form $AdS_d \times B^{n}\times T^{10-d-n}$, with $B^n$ positively curved and with fluxes threading through the compactification, provided the first examples of the holographic AdS/CFT duality \cite{AdS/CFT}.   These can be described in terms of a $d$-dimensional effective potential (as in \cite{Silverstein:2004id}), with a negative curvature-induced term arising from the dimensional reduction of the Einstein term $\sqrt{g}{\cal R}$, played off against a positive term from the flux energy.

In the dual brane construction, these fluxes and the corresponding geometry arise from the presence of color branes (e.g. D3-branes in the canonical $AdS_5 \times S^5$ example and D1-D5 for $AdS_3 \times S^3 \times T^4$) probing the space transverse to their worldvolume directions.  The space probed by these branes takes the form of a cone with base $B^n$, 
\begin{equation}
ds^2=dw^2 + R(w)^2 ds^2_{B}
\end{equation}
with $R(w)=w$.  For our purposes it will be useful to review how this comes about in the following way.  The equations of general relativity
applied to the radius $R(w)$ of the base require
\beq\label{adscone}
\frac{\left(dR/dw\right)^2}{R^2} \sim +\frac{1}{R^2}
\eeq
\noindent
with the $+$ sign corresponding to the positive curvature of $B^{n}$.
This is a radial analogue of the Friedmann equation of cosmology, with $R'/R$ (prime denoting differentiation with respect to the radial coordinate $w$) playing the role of Hubble, and we have included only the curvature term on the right hand side because this is all that contributes in the absence of the color branes.
This has the solution $R = w,$ giving the metric $ds^2 = dw^2 + w^2 ds^2_B$ of a noncompact cone.

In the presence of the color branes, the near horizon $AdS_d$ solution arises from a competition of the positive curvature of $B_n$ against flux terms which must be included on the right hand side of (\ref{adscone})
along with the curvature of the $d$ noncompact dimensions.

Starting from these Freund-Rubin solutions, we will next add ingredients to ``uplift'' the AdS solution to dS, deriving an effective potential which has minima with positive cosmological constant.  Then, we will ask what becomes of the original AdS/CFT brane construction in the process of uplifting.

The method we will use to achieve the uplifting is to introduce, among other things,

\noindent (i) Contributions which overcompensate the positive curvature in the original Freund-Rubin compactification.
One such ingredient is an elliptic fibration of the $T^{10-n-d}$ over $B^n$,
\begin{eqnarray} \label{fibration}
T^{10-n-d} \rightarrow  &{\cal Y}_{10-d}& \nonumber \\
&\downarrow & \nonumber \\ &B^{n}&
\end{eqnarray}
which introduces negative contributions to the scalar curvature that compete with the negative potential term in the original Freund-Rubin compactification \cite{Polchinski:2009ch}.  NS5-branes at real codimension two on the base $B$ also compete with its curvature.  D-branes wrapping all of $B^n$ (along with suitably stabilized anti-branes or other sources canceling their charge) dominate in the expansion in inverse radii and can play an important role in the uplifting, though they are subdominant in the string coupling and hence must be combined with other sources.

\noindent (ii) Orientifolds, at higher codimension than the leading uplifting term, to generate the intermediate negative term in the potential required to obtain a metastable minimum.

We will explain this and related methods in a detailed class of $(A)dS_3$ examples in the remainder of the paper; further examples in four dimensions are in progress \cite{usnext}.  For now, let us assume such a construction exists and analyze its effect on the brane construction and the structure of the resulting holographic dual.

Elliptic fibrations (i) can be thought of as a configuration of 5-branes as in \cite{Greene:1989ya,Hellerman:2002ax}; we will call these ``stringy cosmic 5-branes (SC5s)''.  Since they are extended in the radial direction, they are flavor branes and in general introduce both electric and magnetic matter.  Neveu-Schwarz branes and spacefilling D-branes also contribute flavors. Orientifolds (ii) project the D-brane theory onto a different gauge group, flavor group, and matter content, with unitary groups replaced by orthogonal or symplectic groups.

More significantly, we would like to understand what happens to the space -- the analog of the cone described above -- on which the color branes live.   We will in particular consider what uplifting does to the equation (\ref{adscone}) satisfied by $R$ (the radial modulus of the base) in the absence of the flux contributed by the color branes.  In general, this problem is more complicated than in the simplest AdS/CFT models:  removing the flux will destabilize many moduli in general, leading to radial and/or time evolution of more than just $R$.  In a given construction, one may study this in detail.
However, there is a general qualitative feature of the de Sitter brane construction which follows more simply.

Let us start with a configuration, at some initial time, in which the non-radial moduli are independent of $w$, and carry zero kinetic energy.  We can then focus on the radial modulus $R$, solving its equation of motion by letting it vary radially with $w$.  Given the uplifting, the radial Friedmann equation is now of the form
\beq\label{einsteinds}
\frac{R'(w)^2}{R^2} \sim -\frac{1}{R^{n_1}} + \frac{const}{R^{n_2}}
\eeq
We have taken into account that the positive-curvature term in (\ref{adscone}) has been overcompensated.
We have also included the orientifold stress-energy of the uplifted model, and in order for this to provide an intermediate negative term in the potential we must have $n_1<n_2$.\footnote{For the purposes of the present heuristic discussion, we have not included kinetic mixing of moduli; we will address this below in (\ref{eq:effFRW}) and find that it does not change the qualitative result.}

Since $n_2 > n_1$, there will be points at which $R'(w) = 0,$ so the radius $R$ grows to a maximum size and then proceeds to decrease again.  In the analogue Friedmann equation, this is like a closed universe in the radial direction.  (Note that since we are discussing spatial rather than temporal evolution, the case of a closed universe follows from negative rather than positive curvature.)  The cone of the AdS/CFT brane construction has become a compact space in the de Sitter case.

Now let us add back in the color branes.
In the AdS case, we place color branes at the tip of a cone, and they warp the geometry to produce a Freund-Rubin flux compactification.  In the dS case,
since the radial direction is compact, there is a second tip where $R$ shrinks to zero size.  If we put color branes at one tip and anti-branes at the other, this again generates the flux which plays off against the other ingredients to stabilize the compactification.  The two tips in the brane construction correspond to the two warped throats comprising the de Sitter static patch.  That is, the brane construction corresponding to the uplifted model has automatically produced a microscopic realization of these throats!

\begin{figure}[h]
\begin{center}
\includegraphics[bb = 0 50 600 750, width=10cm]{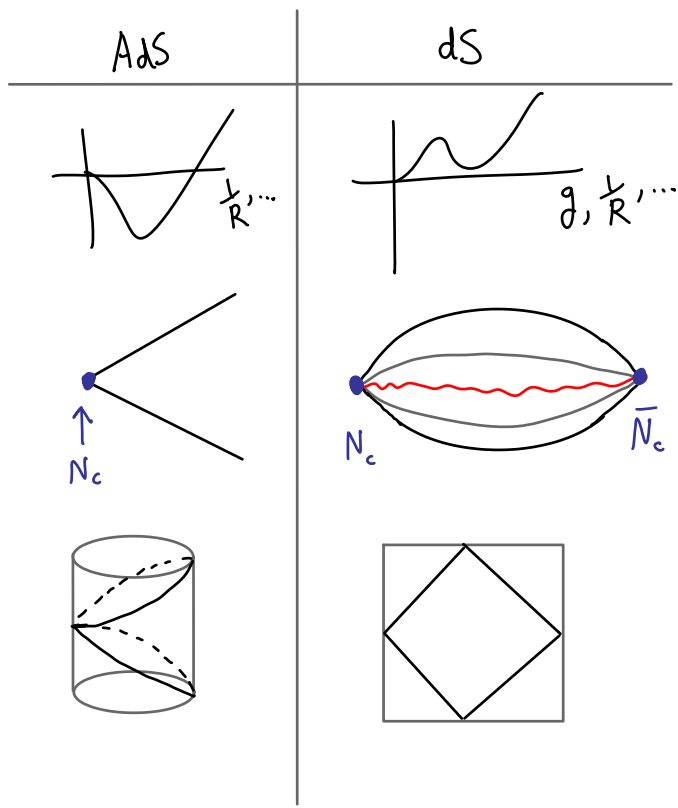}
\end{center}
\caption{de Sitter brane constructions are compact as a result of the net positive potential energy carried by the flavor branes, curvature, and orientifolds.  The two tips with color branes or antibranes correspond to the two warped throats of $dS_d$ in the $dS_{d-1}$ slicing.}\label{fig:braneconstructions}
\end{figure}

As in the cosmological analogue, this geometry can develop curvature singularities at the tips where $R(w)$ shrinks to zero size; these are radial analogues of a big bang and big crunch.  These generalize the conical singularity in familiar AdS/CFT examples.  The nature of the singularities depends on the powers $n_1$ and $n_2$ arising in (\ref{einsteinds}).  The second (orientifold) term dominates the right hand side of (\ref{einsteinds}) near $R=0$. In the case of negative curvature, $n_1=2$ and $n_2>2$; in this case the orientifold term induces a timelike singularity which is worse than conical.  This has to do with the singularity at the cores of the orientifolds, which would be interesting to resolve.  However, in our construction below this question is evaded, as the leading $R$-dependences in (\ref{einsteinds}) will yield $n_1=0,n_2=2$ at fixed string coupling.

In the presence of the flux corresponding to $N_c$ color branes, the right hand side of the radial Friedmann equation (\ref{einsteinds}) acquires one or more additional terms of the form $\sim -N_c^2/R^{2n}$ (with $2n>n_2$).  This dominates at small $R$ and prevents the crunch or bang singularity from happening.  Once all the ingredients are included in a way which yields a complete stabilization mechanism, $R'$ goes to zero for all $w$ (as the moduli are stabilized) and the right hand side of (\ref{einsteinds}) acquires new terms including the $d$-dimensional de Sitter curvature.

In the regime of couplings applicable to the de Sitter solution, the color branes are best described in terms of their dual gravity solution.  The first, simplest examples of AdS/CFT dual pairs had a line of fixed points connecting the regimes of weak and strong coupling in the low energy limit of the brane construction. In cosmological solutions such as this (and also for generic gauge/gravity duals, even some such as \cite{Aharony:1998xz} closely related to the original examples of AdS/CFT), there is not a line of maximally symmetric solutions allowing one to continue between weak and strong couplings regimes of the brane construction.  One may, however, consider weakly coupled, but less symmetric, time dependent backgrounds by analyzing the runaway region near weak coupling and/or large radius.

As we discussed above, the dS/dS correspondence is `semiholographic' in the sense that the Planck mass is finite and $(d-1)$-dimensional gravity does not decouple.  Nevertheless, as we will show below, the dS entropy can be understood parametrically in terms of the degrees of freedom of the brane system.  The reason for this is fairly simple -- the ingredients we add to uplift add a small number of flavors and projections to the original AdS brane system, which does not change its entropy as a function of large quantum numbers such as the dimensions of the color groups.

\section{General techniques}\label{sec:microscopic}

Our technique for stabilizing moduli while uplifting AdS/CFT dual pairs can be thought of as a combination of two familiar methods:  Freund-Rubin stabilization, and the identification of elliptically fibered manifolds with branes on their base.  In this section, we discuss the general methods involved in this construction before coming to an explicit class of examples in \S\ref{sec:dSthree}.

de Sitter model-building in string theory (without a connection to a known holographic dual) has proceeded actively since the discovery of the late-time acceleration of the universe (see e.g.\ \cite{Silverstein:2004id}\ for some reviews with various perspectives on the problem).  Following work anticipating the landscape and its role in interpreting the cosmological constant \cite{Weinberg:1987dv,Bousso:2000xa}, early constructions make use of the positive leading potential from supercriticality \cite{Silverstein:2001xn}\ or from anti-D3-branes \cite{Kachru:2003aw}\ in warped flux compactifications \cite{Giddings:2001yu}\ with non-perturbative contributions to the superpotential.  The latter scenario has provided rich ground for low-energy supersymmetric model building in cosmology and particle physics, but particularly for the goal of understanding de Sitter holography microscopically it may be advantageous to seek simple and explicit de Sitter solutions using perturbative ingredients \cite{Silverstein:2007ac,classicaldS}.

In this line of development, one lesson thus far has been that brane sources with tension $\sim 1/g_s^2$ play a very useful role in de Sitter stabilization; certain no go results follow in their absence.  These are a priori more difficult to control at weak string coupling than D-branes.  One way in which the present work builds further in this direction is to realize such objects via elliptic fibration \cite{Hellerman:2002ax,Polchinski:2009ch}, which incorporates their backreaction.  (In some cases we will find that the core sizes of required solitonic branes are controllably small in any case.)

\subsection{The strategy for stabilization}\label{subsec:eft}

With their backreaction taken into account, the sets of $N_{Dp}$ color branes described above are replaced by corresponding RR fluxes,
\begin{equation}\label{eq:flux1}
\int_{\Sigma^{8-p}} F_{8-p}= N_{Dp}\;\Rightarrow\;F_{8-p} \sim \frac{N_{Dp}}{ \vol(\Sigma^{8-p})}
\end{equation}
where $\vol(\Sigma^{8-p})$ denotes the volume of the surface threaded by the flux. Here we set $\alpha'=1$ and simplify the formulas by omitting numerical factors such as $2\pi$; these will be taken into account in the explicit analysis in \S \ref{sec:dSthree}. Also, the metric signature is taken to be $(- + \ldots +)$.

We look for solutions which are locally of the form
\begin{equation}
dS_d \times B_n \times T^{10-n-d}
\end{equation}
in the presence of background fluxes (\ref{eq:flux1}), plus the flavor branes and orientifolds. The radius of $dS_d$ is denoted by $R_{dS}$. Notice that in general these localized sources will break the isometries of $B_n \times T^{10-n-d}$.

There are two different approaches to this problem. First, one can work directly in ten dimensions, looking for solutions to the equations of motion derived from the (string frame) action
\begin{equation}\label{eq:action}
S =\frac{1}{2\kappa_{10}^2} \int \sqrt{-g^{(10)}}\,\left[ e^{-2\phi}(\mathcal R^{(10)} + 4 (\nabla \phi)^2-\frac{1}{2}|H_3|^2) -\frac{1}{2} |\t F_n|^2\right]+S_{CS} + S_{loc}\,.
\end{equation}
Here $S_{CS}$ denotes the type II Chern-Simons terms, and $S_{loc}$ stands for contributions from localized sources,
\begin{equation}\label{eq:Sloc}
S_{loc} = -  T_p \int \sqrt{-g^{p+1}}\,.
\end{equation}
This method is preferable when practical.  However,
explicit solutions to the equations of motion can be easily obtained only when enough isometries are present, which is not the case here.

Instead, we will analyze the $d$-dimensional effective field theory derived by compactifying (\ref{eq:action}) on $B_n \times T^{10-n-d}$, anticipating a solution with internal dimensions small with respect to the de Sitter radius $R_{dS}$. This requires identifying the light scalar fields which must be stabilized,\footnote{which we will loosely refer to as ``moduli"} computing their effective potential, and finding a minimum with positive cosmological constant. In fact a minimum is not strictly necessary: in order to study accelerated expansion, one requires that any tachyonic masses be small compared to the Hubble scale of the de Sitter solution.  A holographic or semi-holographic description of this situation would be interesting in itself.

To begin with, we will derive an approximate $d$-dimensional moduli potential by averaging the localized sources over the internal space, ignoring the warp factor. Then it must be checked that such a solution can be lifted to a full 10d configuration. The 10d consistency conditions will be discussed in \S \ref{subsec:10dconsistency}\ and addressed in our specific model in \S \ref{sec:dSthree}.

Three of the moduli consist of the dilaton and internal volumes\footnote{We use a tilde on the base size $\tilde R_0$ because in explicit models such as orbifolds, the base may be anisotropic and we will find it useful to reserve the notation $R$ for the curvature radius of the base.}
\begin{equation}\label{eq:vevs}
\tilde R_0^n \equiv \langle \vol(B_n) \rangle\;\;,\;\;L_0^{10-n-d} \equiv \langle \vol(T^{10-n-d}) \rangle.
\end{equation}
In terms of these, the $d$-dimensional Planck scale from dimensional reduction is
\begin{equation}\label{eq:planck}
(M_d)^{d-2} = \frac{\tilde R_0^n L_0^{10-n-d}}{g_{s,0}^2}\,.
\end{equation}
It is useful to introduce fluctuating fields with vanishing VEV,
\begin{equation}
g_s = g_{s,0} e^{ \phi}\;,\;\tilde R = \tilde R_0 e^{\sigma_R}\;,\;L=L_0 e^{\sigma_L}\,.
\end{equation}
At fixed Planck scale, $\phi$, $\sigma_R$ and $\sigma_L$ have kinetic terms independent of the overall volume.

The effective action becomes
\begin{equation}
S_{eff} = M_d^{d-2} \int \sqrt{-g^{(d)}} e^{-2  \phi + n \sigma_R + k \sigma_L}\,\mc R^{(d)} + \ldots\,,
\end{equation}
where here $k=10-n-d$.
The dependence of the Einstein term on the fluctuating scalars is removed by a Weyl rescaling,
\begin{equation}
g^{(d)}_{E\mu\nu} \equiv \left( e^{-2 \phi + n \sigma_R + k \sigma_L}\right)^{2/(d-2)}\,g^{(d)}_{\mu\nu}\,.
\end{equation}
From now on we work in Einstein frame and drop the `E' subindex. Then, the action takes the form
\begin{equation}\label{eq:Seff}
S_{eff}= \int \sqrt{-g^{(d)}}\,\left[M_{d}^{d-2} \left( \mc R^{(d)}- G_{ij}\, g^{\mu \nu} \partial_\mu \sigma^i \partial_\nu \sigma^j\right) -\mc U\right]\,.
\end{equation}
The (positive definite) kinetic term metric $G_{ij}$ for the moduli $\sigma^i$ follows by dimensionally reducing $\int \sqrt{-g^{(10)}} \mc R^{(10)}$ on $B_n \times T^{10-n-d}$ in Einstein frame.\footnote{When backreaction from localized sources is important, a slightly more complicated metric ansatz is required and kinetic terms receive warping corrections. We refer the reader to~\cite{warping} for details. Here we will consistently work in the limit of small warping, where such effects can be ignored.} In our normalization for the moduli, $G_{ij}$ has order one eigenvalues that depend on $d$ and $n$. There is kinetic mixing between $R$ and $L$ ($G_{RL} \neq 0$), reflecting the fact that the overall volume modulus arises from the combination $\tilde R^n L^k$.

The $d$-dimensional Einstein frame potential energy reads
\begin{eqnarray}\label{eq:defU}
\mc U &\equiv& M_d^d \left(\frac{g_s^2}{\tilde R^n L^{10-d-n}} \right)^{d/(d-2)} \left[-\frac{1}{g_s^2}\int \sqrt{g^{(10-d)}}\left(\mc R^{(10-d)}-4(\nabla \phi)^2-\frac{1}{2} |H_3|^2 \right)+ \right.\nonumber\\
&+& \left. \sum_{loc,\;q} T_q\,\vol(\Sigma_{q+1-d})+ \tilde R^n L^{10-n-d} \sum_p \frac{N_{Dp}^2}{\vol(\Sigma^{8-p})^2}\right]\,.
\end{eqnarray}
The first two factors come from the Weyl rescaling and the fact that we work at fixed $d$-dimensional Planck mass $M_d$. The second term inside the square brackets is the contribution from the localized sources (\ref{eq:Sloc}), and $\Sigma_{q+1-d}$ is the cycle wrapped by the q-brane along the internal directions. For D-branes/O-planes, $T_q \sim 1/g_s$; NS5, KK5 and SC5-branes have tension $T \propto 1/g_s^2$ that can compete against curvature if they sit at real codimension two on the base $B_n$.\footnote{The 10d dilaton can vanish or blow up at the cores of localized sources. In our discussion, $g_s$ denotes the $d$-dimensional field, which corresponds to an average value of the 10d mode away from the sources. A similar comment applies to the complex and K\"ahler moduli of $T^{10-n-d}$, which can degenerate at the positions of SC5 branes.} The last term is produced by the flux backreaction eq.~(\ref{eq:flux1}) from the color branes.

So far, we are ignoring contributions from the warp factor derived carefully in \cite{Giddings:2005ff,Douglas:2009zn} which we will address below.
Note that as is standard, with the Weyl rescaling factor in place each term in the effective potential goes to zero at weak coupling or large radius.

\subsection{Stabilization procedure}\label{subsec:abc}

Minima of Eq.~(\ref{eq:defU}) are conveniently analyzed with the ``abc'' method of~\cite{Silverstein:2007ac}, as follows. Before adding the torus fibration, we have curvature
\begin{equation}
\mc R^{(10-d)} \sim \frac{1}{R^2}\,,
\end{equation}
where as mentioned above, the curvature radius $R$ may differ from the
$n^{th}$ root of the volume $\t R$ in anisotropic models such as orbifolds.
The potential energy from positive curvature is
\begin{equation}
\mc U_R \sim - M_d^d\,\left(\frac{g_s^2}{\tilde {R}^n L^{10-n-d}} \right)^{2/(d-2)} \,\frac{1}{R^2}\,.
\end{equation}
The calculations simplify in terms of the variable
\begin{equation}\label{eq:eta}
\eta \equiv \frac{1}{R} \left(\frac{g_s^2}{\tilde R^n L^{10-n-d}} \right)^{1/(d-2)}
\end{equation}
which gives $\mc U_R \sim -M_d^d \eta^2$. We note the useful relation between the Planck scale (\ref{eq:planck}) and the stabilized value of the moduli,
\begin{equation}\label{eq:relation}
M_d = (\eta_0 R_0)^{-1}\,.
\end{equation}

Stringy cosmic branes and NS5 branes give positive contributions to $\eta^2$, competing with and potentially over-cancelling the curvature potential energy if they arise at real codimension two on the base $B_n$. Orientifold planes and D-branes contribute terms of order $\eta^{(d+2)/2}$ with opposite signs; the net effect should give a negative coefficient in front of $\eta^{(d+2)/2}$ (denoted below by $-b(\sigma)$, with $b(\sigma)>0$). Flux energy scales like $\eta^d$ and always gives a positive coefficient $c(\sigma)>0$. Putting everything together, we find an effective potential with the structure
\begin{equation}\label{eq:Uabc}
\mc U = M_d^d\, \eta^2 \left(a(\sigma) - b(\sigma) \eta^{(d-2)/2}+ c(\sigma) \eta^{d-2} \right)\,,
\end{equation}
where here $\sigma^I$ are the moduli different from the combination in Eq.~(\ref{eq:eta}). The functions $a(\sigma)$, $b(\sigma)$ and $c(\sigma)$ are computed from (\ref{eq:defU}).

Let us first consider the AdS case, where only the fluxes (related to color branes) and positive internal curvature are kept:
\begin{equation}\label{eq:cdef}
a(\sigma)=-1\;,\;b(\sigma)=0\;,\;c(\sigma)= R^d \tilde R^n L^{10-n-d} \sum_p \frac{N_{Dp}^2}{\vol(\Sigma^{8-p})^2}\,.
\end{equation}
The $\sigma$ fields are stabilized at the critical points of $c(\sigma)$ (denoted by $\sigma_0$). Plugging this back into eq.~(\ref{eq:Uabc}) gives a minimum
\begin{equation}\label{eq:etaAdS}
\eta_0^{d-2}= \frac{2}{d\, c_0}\;,\;\mc U_0 = -M_d^d\, \frac{d-2}{d} \eta_0^2\
\end{equation}
and a cosmological constant (see eqs.~(\ref{eq:Seff}) and (\ref{eq:relation}))
\begin{equation}
\Lambda_{min}= \frac{\mc U_0}{M_d^{d-2}}=- \frac{d-2}{d} \frac{1}{R_0^2}\,.
\end{equation}
Of course, this is the well-known result that Freund-Rubin solutions supported only by flux and positive curvature have an AdS radius of the same order of magnitude as the internal curvature radius, $R_{AdS}^2\approx R_0^2$.

Moving on to the dS case, the ingredients described above give uplifting terms that set $a(\sigma)>0$, orientifolds plus D-branes to set $b(\sigma)>0$, and flux contributions as in the AdS case. It is instructive to first analyze the background solution in the absence of color branes ($a$ and $b$ nonzero, but $c=0$). This will make contact with the discussion of the radial Friedmann equation (\ref{einsteinds}).

We focus on the radial evolution (coordinate $w$ in the slicing of Eq.~(\ref{dSdSmetric})) of the volume moduli $R$ and $L$. As discussed above, generically some of the moduli will become time dependent; here we restrict to an initial time where the kinetic energy is small compared to the gradient energy from radial variation. Neglecting this time dependence we can extremize the effective action (\ref{eq:Seff}) with respect to $g^{(d)}_{00}$, obtaining
\begin{equation}\label{eq:effFRW}
G_{ij} \nabla^w \sigma^i \nabla_w \sigma^j=\frac{d-2}{d} \mc R^{(d)}-\frac{\mc U}{M_d^{d-2}}\,.
\end{equation}
The left hand side is a positive definite quadratic combination of $R'(w)/R$ and $L'(w)/L$. In general, $R'(w) \neq 0$ sources radial dependence in $L$ through the kinetic mixing. We can solve for $L'(w)/L$ in terms of $R'(w)/R$.  Then using the expression (\ref{eq:defU}) for the potential energy, the right hand side of Eq.~(\ref{eq:effFRW}) has the structure discussed in (\ref{einsteinds}) (after a conformal rescaling that relates 10- and $d$-dimensional Einstein frames). Namely, $\mc U$ has both positive and negative contributions so that the right hand side in (\ref{eq:effFRW}) admits nontrivial roots for $R$. $R(w)$ grows until it reaches this value, and then decreases again.
As discussed in \S \ref{sec:holography}, the effective description reveals that the background space is compact.

Next, placing the color branes and antibranes at the tips $R =0$ gives a nonzero $c(\sigma)$.
There exists a solution with positive energy for
\begin{equation}\label{eq:abc-cond}
1 < \frac{4 ac}{b^2}< \frac{(d+2)^2}{8d}\,,
\end{equation}
evaluated at the minimum of the other moduli $\sigma^I$. The strategy is to first minimize
\begin{equation}
\delta(\sigma) \equiv \frac{4 ac}{b^2}-1
\end{equation}
at a small value, with the potential and minimum then becoming
\begin{equation}\label{eq:etadS}
\mc U = M_d^d\,\eta^2 \left(a \left(1- \frac{b}{2a} \eta^{(d-2)/2} \right)^2 + \frac{b^2}{4a} \delta \,\eta^{d-2}\right) \Rightarrow \eta_0^{(d-2)/2} = \frac{2 a_0}{b_0}\,.
\end{equation}

The positive cosmological constant gives a dS radius
\begin{equation}\label{eq:RdS}
R_{dS}^2 \approx \frac{R_0^2}{\delta_0 a_0}\,.
\end{equation}
Small values of $\delta$ and/or $a$ then lead to solutions with small internal dimensions relative to the de Sitter radius. This was studied for AdS compactifications in~\cite{Polchinski:2009ch}, and will arise in a different way in the examples in the present work.

\subsection{Effects from localized sources}\label{subsec:10dconsistency}

Let us now discuss the ten dimensional consistency of the solutions. Using the dimensionally reduced theory, approximating the sources as smeared, we have explained how the ingredients described above can combine to give a solution of the form $dS_d \times B_n \times T^{10-n-d}$. Now we shall analyze the model from a 10d perspective. We would like to understand under what circumstances there exists a 10d solution to Eq.~(\ref{eq:action}) that, after averaging over the internal space, gives results approximately consistent with the ones derived from (\ref{eq:defU}).

The equations of motion must be solved pointwise in ten dimensions. Some of the ingredients such as O-planes are localized in the internal dimensions; i.e.\ their charge and stress-energy are delta-function supported in some directions.  According to the effective theory (\ref{eq:defU}), these O-planes play off against fluxes and net negative internal curvature to stabilize the moduli.  However, the fluxes and internal negative curvature are not delta-function localized at the positions of the O-planes, and so these effects alone cannot play off of each other pointwise to give 10d solutions.

The missing contributions come from p-forms and warping \cite{Giddings:2005ff,Douglas:2009zn}, which must be consistently included in the effective potential.  As we will shortly review, these effects are small when the sources are dilute enough or have little enough tension that the gravitational and RR potentials they source are small in the bulk of the internal geometry.  In our construction in the next section, this will hold for D-branes and orientifold planes; the elliptic fibration itself does not correspond to well-localized sources, but contributes to the curvature-induced potential energy in a manner we can compute using a sigma model.

Let us discuss explicitly the gravitational backreaction. Gravitational and p-form effects are of the same order of magnitude for BPS objects, so the two analyses are parallel. As argued in e.g.~\cite{Giddings:2005ff, Douglas:2009zn}, the contribution that accounts for the localized stress-energy of the sources is a warp factor $e^A$ multiplying the $(A)dS_d$ metric
which varies over the internal dimensions (as well as conformal factors
in the internal metric, depending on one's conventions for the fiducial internal metric). We will look for solutions with $A \ll 1$ away from the cores of the localized sources.

The equation of motion for the warp factor is of the form
\beq\label{eq:warping2}
\nabla^2 A - (\nabla A)^2 = - \mc R^{(10-d)} +g_s^2 \,|F|^2 + g_s^2\, T^{loc} - g_s^2\, \mc U
\eeq
where we have replaced various powers of $e^A$ by 1, anticipating a solution with $A\ll 1$.
Here $ \mc R^{(10-d)}$ is curvature, $F$ is flux and ${\cal U}$ is the $d$-dimensional effective potential, a constant independent of the internal coordinates.
This equation has the property that for delocalized sources the right hand side would vanish and no nontrivial warp factor would be generated.

The corrections to the effective potential ${\cal U}$ are of order $(\nabla A)^2/g_s^2$.
If $A\ll 1$, and hence $\nabla^2 A \gg (\nabla A)^2$, this means that the corrections are negligible.  That is, the $\nabla^2 A$ term dominates over the $(\nabla A)^2$ term in the equation of motion, providing a mechanism to solve the 10d equations in the presence of the unsmeared, localized sources; while at the same time the correction to the effective potential is of order $(\nabla A)^2$, a subdominant effect. Here we are assuming that no special tuning or cancellations occur among the terms in the effective potential.

As mentioned above, a similar criterion applies to the p-form potential fields. The corrections to the effective potential are of order $ |\nabla C_p|^2$ (as can be read directly from the ten-dimensional action (\ref{eq:action})), while the equations of motion require nonzero $\nabla^2 C_p$.

\section{$dS_3$ worked example}\label{sec:dSthree}

Our $dS_3$ construction builds up from the venerable D1-D5 system \cite{Callan:1996dv} corresponding to an $AdS_3\times S^3/\mathbb Z_k\times T^4$ near-horizon geometry (with the $\mathbb Z_k$ acting freely on the $S^3$). The freedom to take the orbifold order $k$ large will be used to stabilize the internal curvature at a small value. Before analyzing the $dS_3$ construction in detail, we should point out that in our model the internal curvature and string coupling, though consistently small, cannot be taken to be parametrically small. Indeed, specific details of the internal geometry will be found to limit the order of $k$ and the amount of flux that can be turned on in order to get a $dS$ solution. Ongoing constructions in higher dimensions~\cite{usnext} suggest that this is not a general property of our approach.

To summarize the construction:  we will first consider a nontrivial fibration -- allowing the $T^4$ to vary its shape or size over the $S^3/\mathbb Z_k$ -- as in \cite{Polchinski:2009ch,Greene:1989ya,Hellerman:2002ax}.
\begin{eqnarray} \label{fibrationthree}
T^2\times T^2 \rightarrow  &{\cal Y}_7& \nonumber \\
&\downarrow & \nonumber \\ &S^3/\mathbb Z_k&
\end{eqnarray}
This, together with a set of NS5-branes, will contribute positive curvature energy and help to `uplift' the negative potential energy of the $S^3/\mathbb Z_k$.
We will then include orientifolds which produce an intermediate negative term in the potential. Fluxes corresponding to the color D1- and D5-branes contribute a third set of positive terms.

We begin by discussing these ingredients in detail in \S \ref{subsec:ingr}. In \S \ref{subsec:stabilization} we obtain the 3d effective potential and find a $dS_3$ solution with the radii and string coupling self-consistently stabilized. Other modes are included in \S \ref{subsec:D7} and \S \ref{subsec:other-moduli}, and the construction is analyzed from the 10d viewpoint in \S \ref{subsec:localized}. In \S \ref{subsec:entropy} we estimate the de Sitter entropy, and we end in \S \ref{subsec:other-examples} by commenting on other three-dimensional alternative examples.

\subsection{Brane construction}\label{subsec:ingr}

Our construction requires ingredients which are collected in the following table:

\beq\label{braneconstruction} \begin{tabular}{l l l l l  l l l l l l l}

&$0$ &$1$ \vline &$2$ &$3$  &4 &5 &\vline &6 &7 &8 &9\\
  \hline
$D1$  & x   & x  \vline    &      &     &  & &\vline & & & & \\
$D5$  & x   &x  \vline  &     &    &  & &\vline &x &x &x &x \\
\hline
$O5$  &x &x  \vline & &x &x & &\vline & & &x &x \\
$O5'$ &x &x \vline &x & & &x &\vline &x &x & & \\
\hline
$\rho5$ &x &x \vline &x &x & & &\vline & & &x &x \\
$\rho5'$ &x &x \vline & & &x &x &\vline &x &x & & \\
\hline
NS5 &x &x \vline & &x &x & &\vline &x & &x & \\
NS5$'$ &x &x \vline &x & & &x &\vline & &x & &x \\
\hline
$D7$, $\overline{D7}$ &x &x \vline &x &x &x &x &\vline & &x &x & \\
$D7'$, $\overline{D7'}$ &x &x \vline &x &x &x &x &\vline &x & & &x \\
\end{tabular}
\eeq
We will shortly explain each of these ingredients.  First let us describe the underlying geometry.
The $T^4$ lies along the 6789 directions. The 2345 directions correspond to the radial and $S^3/\mathbb Z_k$ directions. As discussed above, the radial direction is compact due to the effects of flavor branes and curvature. The color D1 and D5 branes (and the corresponding antibranes) are then placed at the tips where $S^3/\mathbb Z_k$ vanishes. From the expression for the potential energy below, in the present construction these are conical singularities.

Let us denote $\phi_1=x_2+ix_3$, $\phi_2=x_4+ix_5$.  We can realize
the $S^3$ as a Hopf fibration: $|\phi_1|^2+|\phi_2|^2=R^2$, with the fiber circle along the $\gamma\equiv\arg(\phi_1)+\arg(\phi_2)$ direction.  The base $\mathbb {CP}^1$ of the Hopf fibration is given by gauging this direction.
The $\mathbb Z_k$ orbifold acts as $(\phi_1,\phi_2)\to e^{2\pi i/k}(\phi_1,\phi_2)$, i.e. by a shift on $\gamma$.
The scalar fields we must stabilize include the string coupling $g_s$,  the sizes $R\sqrt{\alpha'}$, $R_f\sqrt{\alpha'}$ of the base and fiber of the $S^3/\mathbb Z_k$, and the size $L\sqrt{\alpha'}$ of the $T^2$ factors in the geometry.  We will address these first.  In general, we must consider all deformations of the 10d fields which are sourced by our ingredients to check if any are unstable.  We will find that axions and anisotropic metric modes are either projected out or are stabilized by the dynamics of our model.

The ingredients are as follows:

$\bullet$  (1) a variation of the K\"ahler moduli $\rho = b_T+i L^2$ of each $T^2$ over the base $\mathbb {CP}^1$. Here $b_T\sim B_{67}L^2=B_{89}L^2$ comes from the 67 and 89 components of the NS-NS two-form potential.

This introduces complex codimension-two branes, i.e. ``stringy cosmic fivebranes" (SC5-branes), as was described in \cite{Hellerman:2002ax}; $\rho$ degenerates at points on the $\mathbb {CP}^1$ corresponding to the positions of these branes.
In general, both the complex structure moduli $\tau$ and the complexified K\"ahler moduli $\rho$ of the $T^2$ fibers could become singular, corresponding to two types of stringy cosmic fivebranes, which we shall call $\tau$5- and $\rho$5-branes respectively.  In the present construction, for simplicity we will use only $\rho5$-branes.

The set of $\rho5$-branes makes two important contributions to the potential energy.
First, recall that a varying complex structure $\tau$ over the base $\mathbb {CP}^1$ subtracts from the scalar curvature.  Since T-duality interchanges $\tau$-fibrations with $\rho$-fibrations, the same holds for the $\rho5$-branes.  A slightly more subtle effect is the following.
Appropriate sets of (p,q) $\rho5$-branes set boundary conditions for $\rho=b_T+iL^2$ at their cores, fixing the size $L$ of (and the axion $b_T$ on) the corresponding $T^2$ fiber to be some value $L_*$ (and $b_*$), usually of string scale. As we introduce other ingredients into our construction, they can cause the (averaged) size $L$ of the $T^2$ fibers to increase, and the variation of $L$ (and $b_T$) results in extra gradient energy of order
\beq\label{rhogradient}
\int_{{\cal Y}_7}d^7y\sqrt g\frac{|\nabla\rho|^2}{g_s^2 \rho_2^2}\sim{\hat n_{\rho}\over{g_s^2 R^2(\alpha')^{3/2}}}{L^4 R^3\over k}\left(\left(\log\frac{L^2}{L_*^2}\right)^2+\frac{(b_T-b_*)^2}{L^4}\right)
\eeq
deriving from the curvature of the fibration.  We write $\hat n_{\rho}$ here to denote the number of stacks of coincident $\rho5$-branes which introduce a boundary condition. This is distinct from the total number of $\rho5$-branes.
We will use the contribution from the gradient energy sourced by the $\rho5$-branes to help stabilize $L$.

To be specific, we will use the following $\rho5$-brane configuration.  Let us describe it in terms of its T-dual, in order to provide a geometrical description in terms of a gauged linear sigma model (GLSM) \cite{Witten:1993yc}.
For reasons which will become clear shortly, we will find it useful to consider branes which locally pin the elliptic fibers at an order one value in string units, using as few branes as possible to accomplish this.
We take a (2,2)-supersymmetric GLSM with charges
\smallskip
\beq\label{chargevectors} \begin{tabular}{ c c c c c c c c c c}
  $\Phi_1$ &$\Phi_2$   &$X_1$ &$Y_1$ &$Z_1$ &$P_1$ &$X_2$ &$Y_2$ &$Z_2$ &$P_2$\\
  \hline
                  0     &0           &2  &3&1 &$-6$ &0 &0 &0 &0 \\
                 0        &0            &0   &0  &0 &0  &2   &3 &1 &$-6$\\
                 $6$   &$6$        & 0 &0 &$-1$ & 0 &0 &0 &$-1$ &0 \\

\end{tabular}
\eeq
under a $U(1)^3$ gauge symmetry.
The D-terms take the form
\begin{eqnarray}\label{modelD} &  \sum_{j=1}^2(2|x_j|^2+3|y_j|^2+|z_j|^2-6|p_j|^2-\ell)^2 \\
\nonumber
& + \left(6|\phi_1|^2+6|\phi_2|^2-|z_1|^2-|z_2|^2 - \xi\right)^2
\end{eqnarray}
Here the Fayet-Iliopoulos parameters correspond to size moduli $\xi\sim R^2$ and $\ell\sim L^2$.

We take a gauge-invariant superpotential of the form
\beq\label{superpot}
\int d^2\theta \sum_{j=1}^2P_j\left\{Y_j^2-X_j^3-Z_j^6 g_j(\phi_1,\phi_2) - X_jZ_j^4f_j(\phi_1,\phi_2)\right\}
\eeq
with $g_1=\phi_1$, $g_2=\phi_2$ of degree 1 and $f_j=0$ in order to respect the gauge invariances (\ref{chargevectors}).
This gives the $T^2$ fibrations as the vanishing loci of Weierstrass polynomials of the form
$$
y_j^2-x_j^3-z_j^6g_j(\phi)-x_jz_j^4f_j(\phi)=0\;,\; {\rm for}\; j=1,2\,.
$$
Each $T^2$ degenerates over a codimension 2 surface, $\phi_1=0$ or $\phi_2=0$ respectively.  These each correspond to
singularities with fixed $\rho\to e^{i\pi/3}$.
Since $f=0$ everywhere, the fibration has constant
$$
\rho=\rho_*=j^{-1}(0)=e^{i\pi/3}
$$
everywhere, not just at these special points where the $\rho5$-branes sit.  In our complete construction, other ingredients will source $\rho$, and deviations from the constant value $\rho_*$ will cost gradient energy (\ref{rhogradient}).

So far, the two $T^2$s vary over a base $\mathbb{CP}^1$ with homogeneous coordinates $\phi_1$ and $\phi_2$.  In this model as it stands, there is a (spacetime supersymmetric) $\mathbb Z_6$ orbifold singularity at $z_1=z_2=0$ descending from the third $U(1)$ gauge symmetry in the table (\ref{chargevectors}).

In our model of interest (\ref{fibrationthree}), the base is not in fact $\mathbb{CP}^1$; it is instead a $\mathbb Z_k$ orbifold of a Hopf fibration over this $\mathbb{CP}^1$, a Lens space.
In the GLSM, the third $U(1)$ gauge transformation parameterizes the Hopf fiber.  For our purposes, we need the model obtained by reducing this continuous gauge identification to a $\mathbb Z_k$ identification.

As in \cite{Polchinski:2009ch} we can obtain the elliptic fibration over $S^3$ (including the Hopf fiber) as the base of a cone.  We introduce the radial direction of this cone along with the Hopf fiber by adding another chiral field $\Phi_0$ to the GLSM,
assigning $\Phi_0$ charge -10 under the third $U(1)$.  (We insist here that the sum of charges cancel, producing a non-compact Calabi-Yau fourfold, in order to preserve SUSY among the $\rho5$-branes.)
In order to incorporate the $\mathbb Z_k$ orbifold we
also mod out by
$$
(\phi_1,\phi_2,z_1,z_2)\to (\alpha^6\phi_1,\alpha^6\phi_2, \alpha^{-1} z_1,\alpha^{-1}z_2)
$$
where $\alpha=e^{2\pi i/k}$ (with the other fields invariant).

From the vanishing of the discriminant $\Delta=27g^2+4 f^3$, this model introduces $n_{\rho}=4$ $\rho5$-branes, significantly fewer than the 24 which would fully cancel the curvature of the $\mathbb{CP}^1$.  This agrees with the fact that beta function for the Fayet-Iliopoulos (FI) parameter $R^2\sim \xi$ describing the size of the base is $(12-2)/12=20/24$ times what it would be in the absence of the nontrivial fibration. (This beta function is proportional to the curvature; in the GLSM it is proportional to the sum of the charges of the fields under the $U(1)$ gauge symmetry corresponding to this FI parameter.)  In this model, the number of defects itself is 2; i.e. $\hat n_{\rho}=2$ in (\ref{rhogradient}).  We will use these numbers for definiteness in our analysis.

Since we do not require the fibration to nearly cancel the curvature, the singularities analyzed in \cite{Polchinski:2009ch} do not arise.  A priori we do not require a hierarchy between the dS and internal curvatures in order to study conceptual questions about de Sitter holography; however, we will obtain such a hierarchy in our explicit construction below for somewhat different reasons from those in \cite{Polchinski:2009ch}.

The additional ingredients are as follows:

$\bullet$  (2) $N_{D5}$ units of RR $F_3$ flux on the $S^3/\mathbb Z_k$, and $N_{D1}$ units of RR $F_7$ flux on ${\mc Y}_7$ (\ref{fibrationthree}).

$\bullet$  (3) An orientifold five-plane wrapped on an unorbifolded $S^1$ in $S^3/\mathbb Z_k$ times one $T^2$. The O5-plane acts as an orientation reversal combined with a reflection on the other $T^2$ and on two of the directions of the $S^3/\mathbb Z_k$. We include a second O5-plane on an orthogonal $S^1$ and the other $T^2$.
Note that the orbifold enhances the effect of the orientifold planes relative to the $\rho5$-branes and other sources which wrap the fiber circle.  In general in de Sitter model building, one needs a negative intermediate term in the potential which competes with the leading term.  This requires the ratio of their coefficients to be large; in our case this ratio is given effectively by $k$.

In terms of the elliptic fibration given above, the orientifold projection acts as $\phi_1 \rightarrow \mp \overline{\phi}_1$, $\phi_2 \rightarrow \pm \overline{\phi}_2$, and $x_j$, $y_j$, $z_j$, and $P_j$ are also projected to their conjugates.

It is interesting to study this effect -- that orientifolds counteract positive energy sources -- in a ten dimensional description.
The O5 metric, at distances large compared to the string scale, looks like
\beq\label{Oplanemet}
ds^2=(1-(\alpha' g_s/r^2))^{-1/2}(-dt^2+dx_\parallel^2) + (1-(\alpha' g_s/r^2))^{1/2}dx_\perp^2
\eeq
The O-planes contract the space around them, more strongly so near the object.
Now consider starting from a metric with a deficit angle induced by a stringy cosmic brane, and orientifold it.  The contraction induced by the O5 will have a more pronounced effect near its core than farther away, since the effect dies off at large $r$ away from the O5.  This reduces the deficit angle.

$\bullet$  (4) An NS5-brane wrapped on an unorbifolded $S^1$ in $S^3/\mathbb Z_k$ and stretched along a one-cycle of each $T^2$. Include another one wrapped on orthogonal directions.  For future convenience, we will take these branes to wrap along the orientifold loci in the base $\mathbb{CP}^1$, which has the effect of reducing the NS5-brane tensions and of projecting out their slippage modes.  An important issue which we will address below is the backreaction of these NS5-branes.  In our ultimate construction below we will ensure that their core sizes are significantly smaller than the base radius.

$\bullet$  (5) Dp-branes and anti-Dp-branes:  to begin with, we will consider a D7-brane wrapped on $S^3/\mathbb Z_k$ and stretched along a one-cycle of each $T^2$, and an anti D7-brane wrapped on the same cycle but in a different discrete Wilson line vacuum. Include also another such pair wrapping the other cycle of each $T^2$.
We put each anti-brane in a nontrivial discrete Wilson line vacuum in order to prevent perturbative brane-antibrane annihilation, as we will explain in \S \ref{subsec:D7}.  In order to decrease the curvature of the base $B_n$, we will find that a simple variant with D5-branes replacing the D7-branes is advantageous.  We will discuss this below in section \S\ref{subsec:fivebranes}\ after working through the D7-brane version of the model.

These ingredients all together break supersymmetry.  However, pairwise many preserve supersymmetry and hence do not attract or repel to leading order.  As in the standard $AdS_3$ model, the D1-D5 color branes are replaced by fluxes in the solution.  In particular, all the other ingredients are pairwise mutually supersymmetric except the $D7$-$\overline{D7}$ pair, whose stability we will explain in detail below.

\subsection{Stabilization mechanism}\label{subsec:stabilization}

To begin with, we will write down a na\"ive 3d effective potential obtained by averaging each source over the compact directions.
This procedure ignores warping which develops as a result of the varying degree of localization of the sources in the internal dimensions \cite{Giddings:2005ff,Douglas:2009zn}.  We will show this to be a self-consistent approximation by analyzing the form of the equations determining the warp factor, finding that the warping necessary to solve the 10d equations of motion contributes a subdominant term to the 3d potential energy (as discussed in general terms in \S\ref{subsec:10dconsistency}).
In essence, we find that the stabilized values of the coupling and inverse radii are small enough to justify our expressions for the stress energy contributed by the various ingredients we have listed.

We denote the radii of the base and fiber of the $S^3/\mathbb Z_k$ in string units by $R$ and $R_f$, so that the volume of $S^3/\mathbb Z_k$ is $2 \pi^2 R^2R_f $.  Below we find $R_f \sim R/k$. Note that the curvature radius of $S^3/\mathbb Z_k$ is $R$. The radii of the $T^4$ are denoted by $L_6$, $L_7$, $L_8$, and $L_9$ with $L_6L_7L_8L_9\equiv L^4$. We also need to consider the field $b_T\sim L^2 B_{67}= L^2 B_{89}$ sourced by the $\rho5$-branes.  Define
\beq
\tilde\eta\equiv{g_s\over {R^2L^2}}\;,\;\;\;\beta\equiv\frac{kR_f}R\,.
\eeq
We find it more convenient to work with the combination $\tilde \eta$ instead of the variable $\eta = k \tilde \eta^2/\beta$ defined in (\ref{eq:eta}). Transforming to $3d$ Einstein frame as in (\ref{eq:defU}), we obtain
\begin{align}\label{potthreed}
{\cal U}\approx &16M_3^3k^3 \left\{\left(4\pi^2-\frac{2\pi^2}{3\beta^2}\left[24-n_{\rho}-\hat n_{\rho}\left(\left(\log\frac{L^2}{L_*^2}\right)^2+\frac{(b_T-b_*)^2}{L^4}\right)\right]+\frac{\pi kn_{NS5}}{L^2\beta^3}\right)\frac{\tilde\eta^4}k\right.\nonumber\\
&\left.-\left( 2\pi R^2-\frac{n_{D7}R^4\beta}{2k}\right){\tilde\eta^5\over\beta^3}+4\pi^2\left(N_{D5}^2 L^4+\frac{(N_{D1}+b_T^2N_{D5})^2}{L^4}+2b_T^2N_{D5}^2\right)\frac{k\tilde\eta^6}{\beta^4}\right\}
\end{align}
Here $M_3$ is the reduced Planck mass.  The first term is the metric flux contribution from the Hopf fibration over $\mathbb{CP}^1$, and the second term (i.e. the square bracket) represents the net curvature introduced by the elliptic fibration (\ref{fibration}), with the presence of $\rho$5-branes. The boundary values $L_*$ and $b_*$ are determined from the GLSM as
$$
\rho_*=b_*+iL_*^2=e^{i\pi/3}\,.
$$
The third contribution to $\tilde \eta^4$ comes from the tension of NS5-branes.

The $\tilde \eta^5$ term receives a negative contribution from the O5-planes, plus a positive term from the D7- and anti D7-branes. For the RR flux contributions we have three terms, the first coming from $|F_3|^2$, the second from $|F_7+\frac12B_2\wedge B_2\wedge F_3|^2$ (which can be understood by T-dualizing the type IIA coupling $|F_4+\frac12B_2\wedge B_2\wedge F_0|^2$ three times), and the final term from $|B_2\wedge F_3|^2$ (since we do not have $F_5$ in our construction).\footnote{We should point out that although na\"ively the zero modes from $B_{67}$ and $B_{89}$ would be projected out by the orientifolds, a nonzero expectation value $b_* \neq 0$ is allowed because $\rho_* = e^{i\pi/3},$ $e^{2\pi i/3}$ are related by a modular transformation. As explained around (\ref{rhogradient}), the fluctuation $b_T\sim L^2 B_{67}=L^2 B_{89}$ away from $b_*$ has nontrivial dependence along the internal directions, so it does not correspond to a zero mode. Physically, this variation is sourced by a competition between $\rho5$ branes and RR fluxes. With these caveats, we will sometimes refer to $b_T$ as an ``axion''; however, it should not be confused with the $B_2$ zero modes analyzed below in \S \ref{subsec:other-moduli}.}

We have included numerical factors such as $2\pi$, according to
$$
T_p =\frac{1}{(2\pi)^p g_s \alpha'^{(p+1)/2}} \;,\; 2\kappa_{10}^2 = (2\pi)^7\alpha'^4\,,
$$
and the quantization of the p-form fluxes
$$
\frac1{(2\pi\sqrt{\alpha'})^{p-1}} \int_{\Sigma_p} F_p \in \mathbb Z\,.
$$
In this expression for the potential, and elsewhere, we have set $\alpha' = 1$ for simplicity.
The various sets of flavor branes including the elliptic fibration, and the orientifolds, are supersymmetric in themselves and also pairwise supersymmetric with each other.  As a result, their contributions to the potential are well approximated by their underlying BPS tension formulae.
Not all factors are known precisely, however; for example the term proportional to $\log(L^2)$ is an approximation of the gradient term (\ref{rhogradient}) by $\nabla^2 \rightarrow \frac{2}{3R^2}$ which we believe to be a reasonable estimate up to factors close to one (based on computations with trial sinusoidal wavefunctions).

We have included the effect of the orientifolds as well as the $\mathbb{Z}_k$ projection in reducing the volume, but we do not know the precise internal geometry taking into account the effects of all the ingredients.
In our best controlled examples below, we will find that starting from the above expression for the potential energy, the curvature in string units, ${\cal R}\alpha'$, comes out to be of order $10^{-3}$ (with other examples giving ${\cal R}\alpha'\sim 10^{-2}$ or $10^{-1}$ depending on the details of the Dp-branes used in the construction).  For this reason, although we will not obtain parametrically large radii, we expect corrections to be reasonably small.
Because we will tune the de Sitter cosmological constant to be somewhat smaller than the internal curvature scale, ${\mc O}(\alpha')$ corrections can affect the depth of the de Sitter minimum. However, since the individual terms in the potential are much larger than this, these effects should only shift the stabilized values of the moduli by a small amount.

At this point it may be useful to emphasize an important distinction between curvature radii and size moduli.  The curvature of our internal dimensions goes like $1/R^2$, but does not get large when the radii $R_f$ and $L$ of the Hopf and elliptic fibrations become small.  These can (and will) be closer to the string scale than $R$ without driving up the curvature and resulting $\alpha'$ corrections.

The potential (\ref{potthreed}) has the form
\beq\label{potabcthreed}
{\cal U}\sim M_3^3(a\tilde\eta^4-b\tilde\eta^5+c\tilde\eta^6)
\eeq
which allows us to use the `abc' technique in \cite{Silverstein:2007ac} to stabilize the moduli. We first minimize $4ac/b^2$ as a function of all other moduli besides $\tilde\eta$ --see discussion around (\ref{eq:abc-cond}). If we can use discrete quantum numbers to tune the minimal value of $4ac/b^2$ to be close to but slightly greater than $1$, the potential (\ref{potabcthreed}) will have a de Sitter minimum with $\tilde\eta$ stabilized near
\begin{equation}
\tilde\eta \approx \frac{2a}{b}\approx \frac{ b}{2c}\,.
\end{equation}

The only $R$ dependence of the potential comes from the coefficient $b$, so we can easily minimize it with respect to $R$ at
\begin{equation}
R^2=\frac{2\pi k}{n_{D7}\beta}\,.
\end{equation}
After that the middle term is reduced to $-2\pi^2k\tilde\eta^5/(n_{D7}\beta^4)$ and $4ac/b^2$ becomes
\begin{align}\label{4acb2threed}
\frac{4ac}{b^2}
=\frac{16n_{D7}^2}{k^2}&\left\{\beta^4-\frac{1}{6}\left[24-n_{\rho}-\hat n_{\rho}\left(\left(\log\frac{L^2}{L_*^2}\right)^2+\frac{(b_T-b_*)^2}{L^4}\right)\right]\beta^2+\frac{kn_{NS5}}{4\pi L^2}\beta\right\}\nonumber\\
&\times\left(N_{D5}^2 L^4+\frac{(N_{D1}+b_T^2N_{D5})^2}{L^4}+2b_T^2N_{D5}^2\right)
\end{align}

Let us focus next on the stabilization of $\beta$. This follows from the factor in curly brackets, which has a three-term structure analogous to (\ref{potabcthreed}):
\beq\label{curly}
\left\{\dots \right\} \equiv \tilde a\beta -\tilde b\beta^2 +\beta^4
\eeq
where
\begin{equation}
\tilde a = \frac{kn_{NS5}}{4 \pi L^2}\;,\;\tilde b = \frac{1}{6}\left[24-n_{\rho}-\hat n_{\rho}\left(\left(\log\frac{L^2}{L_*^2}\right)^2+\frac{(b_T-b_*)^2}{L^4}\right)\right]
\end{equation}
In this case, if we can minimize $\tilde a^2/\tilde b^3$ with respect to $L$ and $b_T$ at
\beq\label{abctilde}
\frac{\tilde a^2}{\tilde b^3} = \frac{4}{27} + \epsilon,
\eeq
with a small positive $\epsilon$ (analogously to $1 \lesssim 4ac/b^2$), then we can minimize $\{\dots\}$ with respect to $\beta$ at a positive small value of $\{\dots\}$,
\beq
\{\dots\} = {3\over 4}\tilde b^2\epsilon\;\;,\;\;\beta = \sqrt{\frac{\tilde b}{3}}\left(1-\frac{9}{8} \epsilon\right)\,.
\eeq
This will in turn help us to tune $4ac/b^2$ to be slightly larger than 1 and will give a parametrically small string coupling.

Minimizing $\tilde a^2/\tilde b^3$ with respect to the axion $b_T$ gives $b_T=b_*$. The RR flux contributions to the potential want to push $b_T$ to 0 but as long as $\epsilon$ is small they are subdominant and only cause a small deviation away from $b_T\approx b_*$. Therefore $\tilde b$ is reduced to $\frac16[24-n_{\rho}-\hat n_{\rho}(\log(L^2/L_*^2))^2]$. Minimizing $\tilde a^2/\tilde b^3$ respect to $L^2$ requires
\beq\label{Lteqn}
24-n_\rho-\hat n_\rho\left(\log\frac{L^2}{L_*^2}\right)^2 = 3 \hat n_\rho \log\frac{L^2}{L_*^2}
\eeq
which we would like to satisfy with a large (though limited) $L$. This relation determines $L$, and then Eq.~(\ref{abctilde}) fixes $k n_{NS5}$,
\begin{equation}\label{eq:kns5}
k n_{NS5}= \frac{4 \pi}{3} \beta \hat n_{\rho} \left(\log \frac{L^2}{L_*^2}\right)L^2+ \mc O(\epsilon)\,.
\end{equation}
Therefore, the order of the orbifold is limited by $L$.

An explicit example of an appropriate $\rho$ fibration was given above, with $n_\rho=4$ and $\hat n_\rho=2$.  With these numbers, we find $L = 2.5$ and $k n_{NS5}=88$; this gives $\epsilon = 0.0016$.

Before going on, let us note that in order to minimize $4ac/b^2$ with respect to $L$ (and requiring (\ref{eq:kns5})), it is enough to take
\beq\label{Lthreed}
L^2=\sqrt{\frac{N_{D1}}{N_{D5}}+b_*^2}.
\eeq
This is also the scaling of the D1-D5 AdS solution. However, this is not strictly necessary, because deviations from this equality produce a tadpole from the flux factor that is suppressed by $\epsilon$ as compared to the mass squared responsible for the first equality. Such a contribution then causes only a small deviation from the solution (\ref{Lteqn}). For simplicity, though, in the formulas below we specialize to the scaling $L^4 \sim N_{D1}/N_{D5}$.

Altogether we obtain (dropping numerical factors)
\beq\label{4acb2finalthreed}
\frac{4ac}{b^2}\sim\frac{n_{D7}^2}{k^2}\left[24-n_{\rho}-\hat{n}_{\rho5}(\log(L^2/L_*^2))^2\right]^2\epsilon N_{D1}N_{D5}\sim 1.
\eeq
Setting $n_{NS5}\sim n_{D7}\sim 1$, Eq.~(\ref{abctilde}) implies
that, parametrically, $k\sim L^2\approx\sqrt{N_{D1}/N_{D5}}$, and (\ref{4acb2finalthreed}) reduces to $N_{D5}\sim1/\sqrt\epsilon$, which gives large $N_{D5}$ and even larger $N_{D1}$. With them it should be possible to tune $4ac/b^2$ to be close to but slightly greater than $1$.
After minimizing $4ac/b^2$ we find a de Sitter minimum at $\tilde\eta\approx2a/b\sim\epsilon/k^2$. The moduli and $dS$ radius scale as follows with the parameters:
\beq\label{scalingsthreed}
R_f\sim {R\over k}, ~~~~ R^2\sim L^2\sim k\sim\sqrt{\frac{N_{D1}}{N_{D5}}}, ~~~~~ N_{D5}\sim\sqrt{\frac1\epsilon}, ~~~~ g_s\sim \epsilon, ~~~~ R_{dS}^2\sim \frac{R^2}{\epsilon}
\eeq
Note that we have obtained large radius and weak string coupling, thanks to the small $\epsilon$. From eq.~(\ref{eq:RdS}), tuning $a \propto \epsilon$ to be small also produces a hierarchy $R_{dS} \gg R$. So our model features parametrically small internal dimensions (compared to the dS scale), giving a gap between 3d moduli and internal KK excitations.

Importantly, the size of $L$ is limited by (\ref{Lteqn}).  The fact that this relation is logarithmic helps drive up the value of $L$, but it is limited by the size of the other terms; as mentioned above we take the minimal values we can obtain for $n_\rho$ and $\hat n_\rho$.

\subsubsection{Numerical example}

We can check this numerically for the potential estimated above, and the results are summarized in the following tables:
\beq
\begin{tabular}{c}
Input data\\
\begin{tabular}{c|c}
\hline
$n_{\rho}$&4\\
$\hat n_{\rho}$&2\\
$\rho_*$&$\exp(i\pi/3)$\\\hline
$n_{NS5}$&2\\
$n_{D7}$&4\\\hline
$k$&44\\
$N_{D1}$&156\\
$N_{D5}$&5
\end{tabular}
\end{tabular}
~~~~~~~~~~~~~~~
\begin{tabular}{c}
Stabilized moduli\\
\begin{tabular}{c|c}
\hline
$R$&9.2\\
$kR_f$&7.5\\\hline
$L$&2.5\\
$b_T$&0.48\\\hline
$g_s$&0.02\\\hline
$\epsilon$&0.002\\
$4ac/b^2$&1.003
\end{tabular}
\end{tabular}
\eeq
Here $n_{NS5} = 2$ and $n_{D7}=4$ are the fewest number of branes required for the setup to have the necessary symmetries in the $T^4$ directions. As can be seen from the above tables, our initial data fix the moduli in a de Sitter minimum. The axion is stabilized at $b_T=0.48$ very close to $b_*=0.5$.

Comparing these numbers to the parametrics above, we see that numerical prefactors break the parametric degeneracy between $L$ and $R$; also, $N_{D5}$ is somewhat smaller than expected, but nevertheless $g_s \ll 1$.  The relation $L^2 = \sqrt{N_{D1}/N_{D5}}$ has not been enforced exactly, but this is a small effect since the contribution of the corresponding tadpole will be suppressed by ${\mc O}(\epsilon)$.
The primary use of the flux quantum numbers $N_{D1}$ and $N_{D5}$ was not to fix $L$, as might be supposed from the form of the effective potential, but to keep $4ac/b^2$ within the allowed range. Interestingly, it is even possible to stabilize all moduli without color fivebranes as long as $\epsilon \ll 1$. In particular we find a dS minimum with $N_{D1}=262$, $N_{D5}=0$, and all other parameters approximately as above.  We will comment further on this possibility when we discuss the scaling of the entropy.

It is worth commenting on the size of $\epsilon$, since taking $ \epsilon \ll 1$ is responsible for achieving a weak string coupling and also boosts the number of degrees of freedom. In particular, in three-term stabilization mechanisms (as opposed to the two-term Freund-Rubin mechanism), there is a priori extra freedom to tune the (A)dS radius large.  Our ability to tune $\epsilon$ small is limited by the size of the large quantum numbers in the problem.  We can express this in terms of $k\sim L^2$ (which is limited by (\ref{Lteqn})).  For example, if we shift $k\to k+1$, we shift (\ref{abctilde}) and hence $\epsilon$ by $\sim k N_{D5}/N_{D1} \sim 1/k$.  Shifting $N_{D1}$ by one would seem to shift $\epsilon$ by an amount of order $1/N_{D1},$ however, this effect is suppressed by $\epsilon$ and the effect of changing $N_{D1}$ is negligible when $\epsilon$ is small.

The mass matrix is positive definite for the input data and the stabilized values found above, and yields masses of order $1/R$ and $\sqrt{\epsilon}/R$.  This follows from the fact that the canonically normalized fields are $\sigma_R\equiv M_3^{1/2} \log R, \sigma_L\equiv M_3^{1/2}\log L, \Phi\equiv M_3^{1/2} \log g_s $
and $\sigma_f\equiv M_3^{1/2} \log R_f$.  Differentiating ${\cal U}$ twice with respect to each $\sigma$ yields contributions of the order of a typical term in ${\cal U}/M_3$ which sources it; this is of order $1/R^2$ for the moduli $\sigma_f$ and $\sigma_L,$ and of order $\epsilon/R$ for $\sigma_R$ and $\Phi$.  To the extent that we tune the de Sitter minima to be smaller than the height of the moduli barriers, these masses are  larger than the de Sitter Hubble scale, and for $\epsilon \ll 1,$ the masses of $\sigma_L,$ $\sigma_f$ are parametrically larger.  For the numbers given above, the smallest of the masses is about one order of magnitude above the Hubble scale.

\subsubsection{Higher order corrections}

Finally, let us consider $\alpha'$ and quantum corrections. Quantum effects are controlled by $g_s^2 \sim \epsilon$, and are further suppressed by the KK scale $1/R^2$ (they have to vanish in the limit in which supersymmetry is restored). These can therefore be safely ignored.

On the other hand, a slightly conservative estimate for the size of the ${\mc O}(\alpha')$ corrections to the GLSM is given by the curvature
\begin{equation}
\alpha' \mathcal R \sim \frac{8}{R^2} \sim 0.1
\end{equation}
in the example above. The factor of 8 comes from relating the $\mathbb{CP}^1$ and $S^3$ radii. This is on the edge of control, since we do not understand all $\mc O(1)$ factors arising from the backreacted geometry and from the gradient energy terms. Such corrections will not affect the moduli stabilization barriers, which are not suppressed by $\epsilon$, but can alter the stabilized value of the Hubble scale.

Strictly speaking, when studying the numerics for the case of small $\epsilon$ we must start with the corrected effective potential to leading order in $\mc O(\alpha')$ and then tune $k$ to find $\epsilon$ small and positive, of order $\gtrsim 1/k$.  We stress once again that the numerics quoted in our example are meant to illustrate the stabilization procedure but are not to be taken as exact. However, metastable de Sitter solutions from our effective potential are quite generic, and we expect the exact solution to be not qualitatively different. Moreover, at the end of the section we present a simple way of pushing the curvature to significantly smaller values, by replacing the flavor D7-branes by flavor D5-branes.

\subsection{$D7$-$\overline{D7}$ stability analysis}\label{subsec:D7}

Let us now elaborate on the stability of the D7- and anti D7-brane pairs, ingredient (5) above. These wrap a $T^2$ fiber and the full $S^3/\mathbb Z_k$. The latter introduces fractional Wilson line vacua.  There are $k$ distinct Wilson line vacua (i.e. non-gauge-equivalent flat connections)
\beq\label{discreteWL}
\left(e^{i\int_{fiber}A}\right)^k=1~~~~\Rightarrow~~~~\int_{fiber}A={2\pi n\over k}~, ~~~ n=0,\dots, k-1.
\eeq
Explicit expressions for these vacua on a Lens space $S^3/\mathbb Z_k$ were given in e.g.\ \cite{WilsonLines}.
Let us put the D7-branes in their $n=0$ vacuum and the anti D7-branes in a vacuum with $n\sim k/2$. We must assess potential instabilities of this configuration from brane-antibrane strings (assessing whether there is a tachyon), and from gauge field modes.  Because the fiber circle is small, for some purposes it is useful to analyze this in a T-dual description.

One can see by periodicity of the gauge field or by T-duality that the size of the circle seen by the Wilson lines is of order \beq\label{Rftilde}
\tilde R_f\sim \frac{1}{R_f}\sim \frac{k}{R} \sim \sqrt{k}
\eeq
where in the last two relations we used our stabilization mechanism (\ref{scalingsthreed}).
This circle being much larger than string scale, the brane and antibrane are separated by a parametrically large distance even if they sit at the same position on the transverse $T^2$.

Next let us analyze the Wilson lines for potential instabilities.  Because the Wilson line vacua are discrete, there is a positive contribution to the mass squared in varying away from the corresponding flat connection.  This scales like the square of the field strength.  Since $F\sim \delta A/R$ ($R$ being the size of the space transverse to the fiber circle), this mass squared goes like $1/R^2$.

There is also a negative contribution to the mass squared from the attraction of the brane and antibrane. This can perhaps be seen and estimated most easily in a T-dual description, with an inverted circle radius (\ref{Rftilde}) and a T-dual string coupling $\tilde g_s \sim g_s/R_f\sim g_s k/R$. The (anti) D7-branes are turned into (anti) D6-branes, wrapping the base $\mathbb{CP}^1$ times the $T^2$ and sitting in diametrically opposite positions on the dual circle. The attractive potential between each brane/antibrane pair is (using the scaling $\tilde R_f\sim\sqrt k\sim R$)
\beq\label{sevensevenpotWL}
{\cal U}_{7\bar 7}\sim-M_3^3\left(\frac{g_s^2}{R^2R_fL^4}\right)^3\left(\tilde g_s^2\times\frac1{\tilde g_s^2}\times R^2L^2\right)\sum_{n_1,n_2,n_3=-\infty}^\infty\frac1{|\vec x-\vec nR|}.
\eeq
The first factor here is the usual Einstein frame conversion factor, the $\tilde g_s^2$ is the 10d Newton's constant in the T-dual frame, the $1/\tilde g_s^2$ is the product of tensions, the $R^2L^2$ is the volume over which the D6-branes are wrapped, and the last factor is the codimension 3 potential. The sum over $\vec n=(n_1,n_2,n_3)$ represents the compactification; we can work on the covering space with a periodic array of localized sources, and then later project by translations in order to compactify.

This potential gives a negative mass squared to the attraction mode between the brane and antibrane pair. Expanding around $x\sim R/2$, we get
\beq
{\cal U}_{7\bar 7}\sim-M_3^3\left(\frac{g_s^2}{R^2R_fL^4}\right)^3\frac{L^2}R\left(x-\frac R2\right)^2.
\eeq
Switching to a canonically normalized kinetic term ${\cal L}_{kin}\sim R^2L^2\dot x^2/\tilde g_s\sim\dot\phi^2$, we get the negative contribution to the mass squared from the attraction of the brane and antibrane pair:
\beq\label{sevensevenMass}
\delta m^2\sim-M_3^3\left(\frac{g_s^2}{R^2R_fL^4}\right)^3\frac{L^2}R\frac{\tilde g_s}{L^2R^2}\sim-\frac{g_s}{R^2}.
\eeq
Although this is parametrically of the same order of magnitude as the Hubble scale (remember that the Hubble scale is tuned), it is parametrically smaller than the positive mass squared arising in deformation away from a flat connection. This mode is therefore perturbatively stable.

\subsection{Stabilizing other moduli}\label{subsec:other-moduli}

So far we have addressed $R$, $R_f$, $L$, $b_T$, $g_s$ and the $D7$-$\overline{D7}$ stability.  In this section we will address the other possibly light directions in scalar field space.  In order to holographically formulate inflationary spacetimes, we must require that all modes be lifted, or if tachyonic that the tachyonic mass be much smaller than the de Sitter Hubble scale.

\subsubsection{Axions}\label{subsubsec:axions}

First, there are potentially light scalars arising as axions from the RR forms and from the B-field. The zero modes surviving the orientifold projection are
\begin{eqnarray}\label{eq:axion-def}
C_2&=& c_1\, dx^6 \wedge dx^7 + c_2\, dx^8 \wedge dx^9\,\\
B_2&=& b_1\, dx^7 \wedge dx^8+b_2 \,dx^6 \wedge dx^9+b_3\, dx^6 \wedge dx^8+b_4\, dx^7 \wedge dx^9\nonumber\,.
\end{eqnarray}
The zero modes from $C_0$, $B_{67}$ and $B_{89}$ are projected out by the orientifold action. (Recall that the field $b_T\sim L^2 B_{67}=L^2 B_{89}$ analyzed before varies along the internal dimensions, so it does not correspond to a zero mode; see discussion around Eq.~(\ref{rhogradient})). Finally, the scalars from $C_4$ threading nontrivial cycles are projected out by the orientifold action.

We have analyzed the dependence of the potential energy of these modes coming from fluxes and from the wrapped D-branes, finding a positive mass matrix for our parameters.  This follows more simply by noting first that the underlying D1-D5 AdS/CFT system has no tachyons (even allowed tachyons) from axions.  The additional ingredients which uplift the system to de Sitter do not render their mass matrix tachyonic.  The orientifolds as just noted project out some modes.  The Dp-branes contribute positive masses to Neveu-Schwarz axions along their worldvolumes; similarly, NS5-branes would contribute positively to the mass squareds for RR axions, though in any case those along the NS5-brane worldvolumes are projected out by the O-planes.  The stabilization of $L$ works slightly differently in the de Sitter case as compared to $AdS$.  This affects the mass matrix for the $b$ axions, but in a way that yields a positive mass squared solution for appropriate values of $N_{D1}/N_{D5}$, including those of our numerical examples.

The $\rho$ fibration contributes a subtle effect lifting $C_2$, as follows.  S-dualizing the $|\tilde F_7|^2$ term described above and integrating by parts gives a term proportional to $|B_2\wedge C_2\wedge F_3|^2$.\footnote{See e.g.~\cite{Cremmer:1998px}. The expressions for $\t F_7$ and $\t H_7$ can be understood by using S- and T-duality and from the anomalous D-brane couplings $S_{WZ}= \int e^B \wedge \sum_p C_p$.}
The combined effect of $F_3$ flux through the base $S^3/\mathbb Z_k$ and $B_2=b_* (dx^6 \wedge dx^7+dx^8 \wedge dx^9)$ produced by the $\rho5$ branes gives positive mass terms to both $c_1$ and $c_2$ in (\ref{eq:axion-def}).

\subsubsection{Moduli of the elliptic fibration and NS5-branes}

The moduli of the elliptic fibration are flat to leading order since they come from superpotential terms in a (2,2) sigma model.  In the explicit model given above with linear polynomials $g^{(j)}(\phi)$, there is only one such deformation.

The NS5-branes wrap contractible cycles in the base $\mathbb{CP}^1$ and could possess slippage modes if they were not wrapped on the orientifold loci in these directions.  In this case the NS5-branes are frozen in place by the orientifold action.  It can be checked that the full solitonic field configuration corresponding to the NS5-brane is compatible with the orientifold action; the compatibility condition is equivalent to the condition that these ingredients be mutually supersymmetric.

\subsubsection{Anisotropic deformations}

The setup (\ref{braneconstruction}) is rather symmetric, and the potential is automatically extremized with respect to directions that break the symmetry.  However, we must ensure that such directions are not too tachyonic.
It is energetically favorable for orientifolds to wrap larger cycles.  Because of this, the O5 and O5$'$ contribute negatively to the mass squared for anisotropies of the tori.
However, the $\rho5$-branes and NS5-branes contribute positively to the mass squared of these modes.  The quantity $4ac/b^2$ (\ref{4acb2threed}) is deformed in the following way by these contributions
\beq\label{4acbtachyons}
\frac{4ac}{b^2} \propto \frac{\epsilon \tilde b^2 + \gamma_1\phi^2}{b_0^2 + \gamma_2\phi^2}
\eeq
where $\gamma_1$ and $\gamma_2$ are positive quantities which do not scale down with $\epsilon$.  The $\gamma_2$ term comes from the orientifold.  The tachyonic mass squared that it imparts is suppressed by the power of $\epsilon$ in the numerator.  There is no such suppression of the positive mass squared from the $\rho5$s or NS5s.  So the net effect is a positive mass squared for these anisotropic modes.\footnote{The mode $L^2_{78} \neq L^2_{69}$ is a special case: the only contribution is a positive mass term from D7-branes. The energy of the other ingredients (NS5s, $\rho5$s and orientifolds) is independent of this field. In the case where the D7-branes are replaced by flavor D5s wrapping $S^3/\mathbb{Z}_k$, this direction becomes flat at the classical level.}

However, anisotropies of the base are more subtle.  The O-planes that we have prescribed can elongate without breaking any of the symmetry.  To see this, we can coordinatize the $S^3$ as follows.  Set $\tilde\phi_1\equiv x_3+ix_4\equiv \rho_1e^{i\gamma_1},\tilde\phi_2\equiv x_2+ix_5\equiv \rho_2e^{i\gamma_2}$.
The round $S^3$ metric is
\beq\label{metRfour}
ds^2 = |d\tilde\phi_1|^2 + |d\tilde\phi_2|^2 = d\rho_1^2 + \rho_1^2d\gamma_1^2 + d\rho_2^2+\rho_2^2d\gamma_2^2
\eeq
with $\rho_1^2+\rho_2^2=R^2$.  Using this latter relation, we can set $\rho_1=R \sin\kappa, \rho_2=R\cos\kappa$ with $0\le\kappa\le \pi/2$.  This gives metric
\beq\label{metRfourII}
ds^2=d\kappa^2+\sin^2\kappa d\gamma_1^2 + \cos^2\kappa d\gamma_2^2
\eeq
The O5 lies along $\gamma_1$ at $\kappa=0$, while the O5$'$ lies along $\gamma_2$ at $\kappa=\pi/2$.
It is possible to shrink the $\kappa$ direction without breaking any of the symmetry, elongating the $\gamma$ directions to maintain constant volume.  Therefore, the O-planes alone produce a tadpole in this direction.

However, because of the tuning down of $\epsilon$ above, the NS5-branes and $\rho5$-branes each contribute more to the forces in the problem. Indeed, while the O-planes compete against the net term $\{\dots\}\propto\epsilon\ll 1$ above (\ref{curly}), NS5-branes and $\rho5$-branes contribute larger, individual contributions to $\{\dots\}$.
The NS5-branes push oppositely to the orientifold planes. The $\rho5$-branes stretch in the $\kappa$ direction, so the combined effects of the $\rho5$-branes and NS5-branes together stabilize this direction.

\subsection{Localization of sources and the warp factor}\label{subsec:localized}

After having found a consistent solution in the 3d effective theory, we should analyze the ten dimensional consistency of the construction.
Following the discussion in \S \ref{subsec:10dconsistency},
let us argue that the localization of the sources will not appreciably change the potential (\ref{potthreed}), though it requires taking into account a slowly varying warp factor.

In the effective 3d theory (\ref{potthreed}) the localized sources (for instance O5 planes) play off against fluxes and net negative internal curvature to stabilize the moduli. Since these are not delta-function localized at the positions of the O5-planes, solving the 10d EOMs pointwise requires a warp factor $e^A$ multiplying the $(A)dS_3$ metric
which varies over the internal dimensions~\cite{Giddings:2005ff,Douglas:2009zn}:
\beq\label{warpedadsmetric}
ds^2 = e^{2A(y)}ds^2_{AdS_3} + e^{-2A(y)}\tilde{g}_{ij}dy^i dy^j
\eeq

Recall from the discussion around (\ref{eq:warping2}) that if the equation for the warp factor can be solved (by having nonzero $\nabla^2 A$) with $A \ll 1$, then the corrections to the effective potential are negligible since they are of order $(\nabla A)^2/\kappa^2$. This provides a mechanism for solving the 10d equations in the presence of localized sources while keeping corrections to the potential subdominant. Therefore, let us check that in the present construction the condition $A\ll 1$ holds away from the cores of the localized sources (whose tensions already take the cores into account).


First consider the O5-planes as a source for $A$.  Note that the fiber circle being small, these effectively wrap a T-dual circle of size $\tilde R_f\sim 1/R_f$; there are three directions $\vec y_\perp$ transverse to the O5s.  Schematically we have
\beq\label{warpeqn}
\nabla^2 A(\vec y_\perp) \sim \frac{1}{g_s}\frac{1}{R_f}g_s^2 \sum_{n_R,n_1,n_2}\delta^{(3)}(\vec y_\perp-\vec y_0-\vec n R)  +{\rm other ~ sources},
\eeq
the first factor here being the $1/g_s$ tension, the second the wrapped T-dual circle, and the $g_s^2$ factor arising from Newton's constant $\kappa^2$.  Once again, we can work on the covering space with a periodic array of localized sources, and then later project by translations in order to compactify.  Let us look at $A$ at a point $\vec y_\perp$ halfway in the middle of the localized O5 sources.  This gives us
\begin{align}
A\left(\frac{R}{2}, \frac{L}{2}, \frac{L}{2}\right) & \sim g_s\frac{1}{R_f} \sum_{n_R,n_1,n_2}\frac1{\sqrt{R^2\left(n_R-\frac{1}{2}\right)^2+L^2\left(n_1-\frac{1}{2}\right)^2+L^2\left(n_2-\frac{1}{2}\right)^2}} \nonumber\\
&+{\rm contributions ~ from ~ other ~ sources.}
\end{align}
In the covering space, the sum on $\vec n$ represents the contributions from localized sources farther and farther away from the point $\vec y_\perp$.  As discussed above, if all sources were smeared, the gradient of $A$ would cancel.
So the full expression for $A$, including all localized and homogeneous sources, is essentially the difference between the sum on $\vec n$ and the integral over $\vec n$.  Since $L$ and $R$ are of the same order of magnitude, the overall contribution to the warp factor is $A \sim \mc O(g_s)$, and the localization of the source can be ignored.
Similarly, the gravitational potential sourced by the D7-branes is suppressed by $g_s$, leading to $A\ll 1$ when combined with the full complement of sources.

The $\rho5$-branes would produce $A\sim 1$ in a similar way if we treated them as putatively localized sources of stress-energy.   Indeed, they do have long-range effects on the geometry; they correspond to a nontrivial fibration (\ref{fibration}) described by a Weierstrass model.  The corrected target space is described by the infrared regime of the gauged linear sigma model (GLSM) presented above.  Their contribution to the curvature and hence to the moduli potential is of the same form as it would be in the na\"ive estimate based on the tension of an isolated brane.  The most familiar example of this is the case of elliptically fibered Calabi-Yau manifolds, where the fibration exactly cancels the curvature of the base.  This occurs for example in the case of 24 stringy cosmic fivebranes on a base $\mathbb{CP}^1$.  More generally, the GLSM beta function $\sim 24$ for the running of the FI parameter $\xi$ corresponding to the size $R^2$ of the base is shifted by the elliptic fibration to $\sim (24-n_\rho)$.

We expect similar statements to hold for the NS5-branes (ingredient (4) above).  By themselves, these NS5-branes could be described including backreaction in a half-flat approximation by recognizing them \`a la \cite{Hellerman:2002ax} as a different set of $\rho5$-branes with monodromy $\rho\to \rho+1$.  These are holomorphic with respect to a different choice of complex coordinates on the $\mathbb{R}^4\times T^4$ of our brane construction, so it is not trivial to describe both sets of $\rho5$-branes using the same GLSM.  However, the two sets are mutually supersymmetric and their BPS tensions combine additively. (They are U-duals of D-brane combinations with 4 Neumann-Dirichlet directions; such combinations have no binding energy.)  For this reason we expect their contributions to be well described by the tensions we included in the moduli potential (\ref{potthreed}), at least in cases where their core size is sufficiently small.  In the D7-brane model so far considered, this is marginal; the core size of the NS5 is not smaller than $R,$ as we will discuss below in \S \ref{subsec:fivebranes}.  However, the variants in \S \ref{subsec:fivebranes}\ involving D5-branes will push the parameters to where the NS5-brane has a string-scale core, much smaller than $R$.  It would be useful to develop techniques to simultaneously control multiple types of $1/g_s^2$-tension branes more explicitly.

\subsection{Entropy and brane construction}\label{subsec:entropy}

Given the above scalings (\ref{scalingsthreed}), the entropy scales parametrically like
\beq\label{entropythreed}
{\cal S}\sim M_3 R_{dS}\sim \frac{L^4 R^4}{k g_s^2\epsilon^{1/2}}\sim \frac{kN_{D1}N_{D5}}{\epsilon^{3/2}}
\eeq
This scaling of the entropy is the same as in the corresponding AdS model when $\epsilon<0$.
This makes sense, since the $\rho5$-brane flavors and the orientifold projections are only of order 1 in number, so that the scaling is as in the corresponding D1-D5 AdS orbifold quiver theory up to an enhancement by $\epsilon^{-3/2}$. This factor can be understood as in section 3.5 of \cite{Polchinski:2009ch}: pull out a color brane from a tip of the cone, and count the number of ways of winding strings around the base of the cone up to an energy cutoff comparable to the energy of a string stretching radially to the tip.  This gives an enhancement factor $\sim (R_{dS}/R)^3\sim \epsilon^{-3/2}$.
As a result, we can read off the parametric scaling of the Gibbons-Hawking entropy from the D1-D5 system.

Many details come into the precise coefficient over which we have no control at present.
As in \cite{Polchinski:2009ch}, our main handle on the holographic dual is through its brane construction.  It would be interesting to develop tools to study this theory in more detail.\footnote{An intriguing possibility is that as in \cite{Kachru:2009kg}, the holographic duals may only exist as cutoff theories.}  With the brane construction in hand, however, one can immediately study further questions of interest such as the microscopic description of decays out of the metastable dS vacuum and their holographic description.\footnote{Other macroscopic proposals for de Sitter holography which are based on different ways of slicing the spacetime include \cite{dSCFT,hats}.  It would be interesting to study whether our dS construction building from AdS/CFT might also provide a microscopic realization of these ideas.}

As mentioned in \S2, this model has a brane construction for which the radially evolving $R$, at fixed initial values of the other moduli, solves the equation (\ref{einsteinds})
\beq\label{einsteindsexample}
\frac{R'(w)^2}{R^2} \sim -1 + \frac{const}{R^{2}}
\eeq
for which the singularities at the tips $R\to 0$ are conical.  The first term here comes from the D7-branes and anti D7-branes, and the second includes the O5-planes which are at codimension two on the base.

It is also possible to stabilize the construction with $N_{D5}=0,$ as mentioned ealier.  In this case the parametric scaling is
\beq\label{entropynoF3}
{\cal S}\sim M_3 R_{dS}\sim \frac{L^4 R^4}{k g_s^2\epsilon^{1/2}}\sim \frac{N_{D1}^2}{k\epsilon^{3/2}}
\eeq
In this case the tuning of $\epsilon$ plays the dominant role.  It would be interesting to understand this case better as it does not appear to arise directly from a known AdS/CFT dual pair, as is the case for most models in the landscape.



\subsection{Alternative examples}\label{subsec:fivebranes}\label{subsec:other-examples}

In this subsection we provide two alternative examples in which the radius of curvature $R$ is pushed to larger values and we therefore have better control of $\alpha'$ corrections.

First, we can replace the D7- and anti D7-branes with a single pair of D5- and anti D5-branes wrapping the $S^3/\mathbb Z_k$, and put them in different Wilson line vacua to prevent perturbative brane-antibrane annihilation:
\beq\label{fivebraneconstruction} \begin{tabular}{l l l l l  l l l l l l l}
&$0$ &$1$ \vline &$2$ &$3$  &4 &5 &\vline &6 &7 &8 &9\\
  \hline
$D5$, $\overline{D5}$  & x   & x  \vline    &x      &x     &x  &x &\vline & & & & \\
\end{tabular}\eeq
The stabilization proceeds as before with the middle term $-b\tilde\eta^5$ in the effective potential replaced by
\beq\label{potfivebrane}
\frac{\mathcal U}{M_3^3} \supset - 16 k^3\left(2\pi R^2-\frac{n_{D5}R^4\beta}{2k L^2}\right)\frac{\tilde\eta^5}{\beta^3}\,.
\eeq
This gives
\begin{equation}
R^2 = \frac{2\pi k L^2}{n_{D5} \beta}\,,
\end{equation}
and the rest of the stabilization proceeds as before. Note that $n_{D5}$ is the number of flavor D5- and anti D5-branes as specified above, which should be distinguished from $N_{D5}$, the number of color D5-branes wrapping different directions in our brane construction. With $n_{D5}=2$, we find $R \approx 33$ (as compared to $R \approx 9$ in the model with flavor D7 branes); now $\alpha'$ corrections are of order $8/R^2=7 \times 10^{-3}$.

We can push $R$ to even larger values by wrapping the flavor D5-branes on an unorbifolded $S^1$ in the $S^3/\mathbb Z_k$, the fiber $S^1/\mathbb Z_k$, and one of the $T^4$ directions:
\beq\label{fivebraneconstruction2nd} \begin{tabular}{l l l l l  l l l l l l l}
&$0$ &$1$ \vline &$2$ &$3$  &4 &5 &\vline &6 &7 &8 &9\\
  \hline
$D5$  & x   & x  \vline    &(x)      &\ x     &\ x  &(x) &\vline &x & & & \\
$D5'$  & x   & x  \vline    &\ x      &(x)     &(x)  &\ x &\vline & &x & & \\
\end{tabular}\eeq
Here the parentheses indicate that the true D5-brane locus is a combination of these dimensions.  Note that we do not need to include anti D5-branes here because the D5-branes are wrapping a contractible cycle in the $S^3/\mathbb Z_k$, but in order to stabilize the anisotropic mode $L^2_{78}\neq L^2_{69}$ we need two D5-branes wrapping the 6 and 7 directions respectively. Again, the middle term of the effective potential is replaced by
\beq\label{potfivebrane2nd}
\frac{\mathcal U}{M_3^3} \supset - 16 k^3 \left(2\pi R^2-\frac{\pi n_{D5}R^3\beta}{k L}\right)\frac{\tilde\eta^5}{\beta^3}
\eeq
which stabilizes $R=4kL/(3n_{D5}\beta)$. With $n_{D5}=2$, the value of $R$ is further increased from 33 to 91, with the $\alpha'$ corrections of order $8/R^2=1 \times 10^{-3}$.

Another advantage of these two alternative constructions is that the size of the NS5-brane core is much smaller than the radius $R$. A simple way of looking at this is to take the T-dual along the fiber direction $R_f$: the NS5-branes are turned into KK5-branes with a fiber size $\tilde R_f=1/R_f=k/(\beta R)$. In our previous construction with D7-branes, this size is of the same order of magnitude as $R$, as we can see from the parametric scaling $R^2\sim k$. However, if we replace the D7-branes with D5-branes as in the first example above, we have $R^2\sim kL^2\sim k^2$ and the fiber size $\tilde R_f\sim k/R\sim 1$ is much smaller than $R$. In the second example above, we have $R^2\sim k^2L^2\sim k^3$ and the fiber size $\tilde R_f\sim k/R\sim1/\sqrt k$ is again much smaller than $R$. This is also confirmed by the numerics: with the previous D7 construction we have $\tilde R_f=k/(\beta R)=5.9$ not much smaller than $R=9.2$, with the first D5 construction we have $\tilde R_f=1.64$ much smaller than $R=33$, and with the second D5 construction we have $\tilde R_f=0.59$ much smaller than $R=91$.

The $D7$-$\overline{D7}$ stability analysis can be repeated in the first D5 example, with analogous results: there is a negative mass squared of Hubble scale, but the deformations away from the flat connection give large positive contributions to the mass squared and keep the mode perturbatively stable.  In the second example, the D5-branes are mutually supersymmetric but may possess slippage modes because they wrap contractible cycles on the base $\mathbb{CP}^1$; placing the branes at the orientifold loci (the effect on their tension is already included in (\ref{potfivebrane2nd})) projects these modes out and freezes the branes in place.

\section{Discussion and Future Directions}\label{sec:discussion}

Similar methods apply to the construction of semi-holographic de Sitter models in other dimensions.  Work on four dimensional examples of this kind is in progress \cite{usnext}.  A promising class of candidate models arises from  type IIA string theory on an orientifold of an elliptic fibration over $B_4$ (e.g. $\mathbb{CP}^2$ or $\mathbb{CP}^1\times\mathbb{CP}^1$) which over-compensates the curvature energy of $B_4$, along with RR 2-form and 6-form flux as well as additional flavor branes.

This work raises many interesting questions about the nature of the dual implied by the brane constructions.
Let us make a few comments here.
The compactness of the brane construction implies propagating $d-1$-dimensional gravity.  As we have discussed, the fact that this is coupled to a matter sector with large entropy makes for a useful albeit semi-holographic duality; $d-1$-dimensional gravity is weakly interacting over a large range of scales because of the enhancement of the $d-1$-dimensional Planck mass induced by the large number of species.  (In the $d=3$ case studied explicitly in this
paper, the $d-1$ dual is Liouville gravity coupled to a large-c matter sector, as anticipated macroscopically in \cite{Alishahiha:2004md}\ and proposed for somewhat different reasons in \cite{hats}.)  By the same token many other modes are dynamical.  In particular flavor groups are dynamical in $d-1$ dimensions, i.e.\ the flavor symmetry is weakly gauged; the flavor groups have a large number $\sim N_c$ of matter fields charged under them which screen their interactions.  This is analogous to the weak dynamical gravity in the $d-1$ dimensional description, with large-$c$ matter sector.

One basic question is whether the $d-1$-dimensional matter theory here is UV complete by itself (and just happens to be coupled to gravity in the case that it arises as part of a dual for de Sitter).  Another possibility, analogous to the situation obtained for non-supersymmetric warped throats in \cite{Kachru:2009kg}, is that the matter theory only exists as a low energy effective theory.  In general, we would like to understand the couplings of the matter degrees of freedom to Liouville gravity in our setup.

A related question concerns the microscopic interpretation of the entropy.  So far we obtained a parametric result, but not the precise coefficient.  Even in AdS/CFT, obtaining the precise coefficient is difficult; for example in the $\mathcal{N}=4$ supersymmetric Yang Mills theory there is a famous ratio of 3/4 between the strong and weak coupling results.
In the present case, the coupling to $d-1$-dimensional gravity is a further complication.  In particular, altogether the central charge vanishes in a theory of gravity, something borne out by the macroscopic calculations in \cite{Alishahiha:2004md}.  Perhaps in the case $d-1=2$, the {\it effective} central charge is the appropriate notion of a count of degrees of freedom; this is sensitive to the matter contribution.

Our construction is reminiscent of the attempt \cite{Danielsson:2001xe}\ to count Schwarzschild entropy with a brane-antibrane system.   In the present case, there is a variety of brane constructions which reflects the landscape of possible solutions:  each de Sitter solution corresponds to a definite brane construction.

Our solutions eventually decay.  Decay modes include the strongly coupled, exponentially suppressed version of brane-antibrane annihilation: Schwinger decay of the flux dual to the color branes.  The decay of de Sitter solutions into each other or to regions of zero or negative cosmological constant is central in many attempts to formulate the landscape \emph{en masse}, and concrete input from microscopic brane constructions may be useful.

\subsection*{Acknowledgements}
We are grateful to Joe Polchinski for many interesting discussions and comments on a draft.  We also would like to  thank M. Douglas, D. Morrison and S. Shenker for extensive discussions of various aspects of this work, and R. Flauger, B. Freivogel, S. Kachru, R. Kallosh, L. Susskind, D. Tong, and E. Witten for useful comments.
This work was supported by the National Science Foundation under grant PHY05-51164, the UCSB Department of Physics, the DOE under contract DE-AC03-76SF00515, and by a William K. Bowes Jr.\ Stanford Graduate Fellowship.


\begingroup\raggedright\endgroup

\end{document}